\documentclass[aps,pre,twocolumn]{revtex4}
\usepackage{macros2erev4} 
\usepackage{amssymb,graphicx}
\usepackage{times}
\newcommand{\nn}{\nonumber}

\newcommand{\bilderscale}{0.35}

\newif\ifpdf
    \ifx\pdfoutput\undefined
    \pdffalse 
    \newcommand{\fig}[2]{\includegraphics[width=#1]{./figures/#2.eps}}
    \newcommand{\Fig}[1]{\includegraphics[width=\columnwidth]{./figures/#1.eps}}
        \newlength{\bilderlength}
    \newcommand{\usebilderscale}{\bilderscale}
    \newcommand{\bilderskip}{\hspace*{0.8ex}}
    
    \newcommand{\diagram}[1]{%
    \settowidth{\bilderlength}{\bilderskip%
    \includegraphics[scale=\usebilderscale]{./figures/#1.eps}\bilderskip}%
    \parbox{\bilderlength}{\bilderskip%
    \includegraphics[scale=\usebilderscale]{./figures/#1.eps}\bilderskip}}
    
\else
    \pdfoutput=1 
    \newcommand{\fig}[2]{\includegraphics[width=#1]{./figures/#2.pdf}}
    \newcommand{\Fig}[1]{\includegraphics[width=\columnwidth]{./figures/#1.pdf}}
    \newlength{\bilderlength}
    \newcommand{\usebilderscale}{\bilderscale}
    \newcommand{\bilderskip}{\hspace*{0.8ex}}
    
    \newcommand{\diagram}[1]{%
    \settowidth{\bilderlength}{\bilderskip%
    \includegraphics[scale=\usebilderscale]{./figures/#1.pdf}\bilderskip}%
    \parbox{\bilderlength}{\bilderskip%
    \includegraphics[scale=\usebilderscale]{./figures/#1.pdf}\bilderskip}}
    
    \pdftrue
\fi
\newcommand{\sgn}{{\mathrm{sgn}}}
\newcommand{\rme}{{\mathrm{e}}}
\newcommand{\rmd}{{\mathrm{d}}}

\begin{document}
\title{\sffamily
\bfseries
\Large Higher correlations, universal
distributions and finite size scaling in the field theory of depinning}
\author{\sffamily\bfseries\normalsize Pierre Le Doussal{$^1$} and Kay
J\"org Wiese{$^2$} \vspace*{3mm}}
\affiliation{{$^1$} CNRS-Laboratoire de Physique Th{\'e}orique de
l'Ecole Normale Sup{\'e}rieure,
24 rue Lhomond, 75005 Paris,  France.\\
{$^2$} KITP, University of
California at
Santa Barbara, Santa Barbara, CA 93106-4030, USA\medskip }
\date{\small\today}
\begin{abstract}
Recently we constructed a renormalizable field theory up to two loops
for the quasi-static depinning of elastic manifolds in a disordered
environment. Here we explore further properties of the theory. We show
how higher correlation functions of the displacement field can be
computed. Drastic simplifications occur, unveiling much simpler
diagrammatic rules than anticipated. This is applied to the universal
scaled width-distribution. The expansion in $d=4 - \epsilon$ predicts
that the scaled distribution coincides to the lowest orders with the
one for a Gaussian theory with propagator $G(q)=1/q^{d+2 \zeta}$,
$\zeta$ being the roughness exponent. The deviations from this
Gaussian result are small and involve higher correlation functions,
which are computed here for different boundary conditions.  Other
universal quantities are defined and evaluated: We perform a general
analysis of the stability of the fixed point. We find that the
correction-to-scaling exponent is $\omega =-\epsilon$ and not
$-\epsilon /3$ as used in the analysis of some simulations.  A more
detailed study of the upper critical dimension is given, where the
roughness of interfaces grows as a power of a logarithm instead of a
pure power.
\end{abstract}
\maketitle


\section{Introduction} Understanding the behavior of an elastic
interface in a random potential is important for many experimental
systems and still offers a considerable theoretical challenge
\cite{BookYoung,Kardar1997,DSFisher1998,BlatterFeigelmanGeshkenbeinLarkinVinokur1994}.
It is expected that below the upper critical dimension
$d_{\mathrm{uc}}$ the interface is pinned by arbitrarily weak
disorder, into some rough configurations and that at zero temperature
it can acquire a non-zero velocity under an applied force $f$ only if
$f$ is larger than the depinning threshold $f_c$. A functional
renormalization group (FRG) method predicts that $d_{\mathrm{uc}}=4$
for the statics \cite{DSFisher1986}, and for the simplest universality
class, the so called isotropic depinning
\cite{NattermannStepanowTangLeschhorn1992,NarayanDSFisher1992b,NarayanDSFisher1993a}.

There has been recent progress towards a precise description of the
depinning transition. From the theory side, the FRG for single
component manifolds, originally studied to one loop in an expansion in
$\epsilon=d_{\mathrm{uc}}-d$, has now been extended to a field theory
shown to be renormalizable to two loops. Renormalizable, we recall,
means it has a well defined continuum limit, which is independent of
all microscopic details, and thus ensures universality of large scale
observables. Presumably there exists a fully renormalizable theory to
all orders, with full predictive power
\cite{ChauveLeDoussalWiese2000a,LeDoussalWieseChauve2002,LeDoussalWieseChauve2002a}. From
the side of numerics a novel powerful algorithm allows to obtain the
configurations at (or just below) depinning with much improved
accuracy \cite{RossoKrauth2001a,RossoKrauth2001b,RossoKrauth2002}.  A
reasonable agreement between the two methods was found in a
measurement of the roughness exponent $\zeta$, especially the clear
conclusion that $\zeta > \epsilon/3$ contrarily to a previous
conjecture \cite{NarayanDSFisher1992b,NarayanDSFisher1993a}
($\zeta=\epsilon/3$) based on the 1-loop analysis.

The field theory of depinning in its present form is unconventional in
that one must work with a non-analytic action.  This peculiar feature
is a deep part of the physics of the problem and necessary to avoid
the so called dimensional reduction. It makes the perturbation theory
superficially ``ambiguous''. A non-trivial step taken in
\cite{ChauveLeDoussalWiese2000a,LeDoussalWieseChauve2002,LeDoussalWieseChauve2002a}
to define the theory at $T=0$ as the limit $v\to 0$ of the moving
phase, was to assume that the interface-position is monotonic in
time. This removes the ambiguity and, remarkably, leads to a
renormalizable theory, to at least two loops
\cite{ChauveLeDoussalWiese2000a,LeDoussalWieseChauve2002,LeDoussalWieseChauve2002a}.
This is supported by the ``non-crossing theorems'' which apply to
single component depinning and, remarkably, is the same property
allowing to show ergodicity and to construct an efficient algorithm to
find the exact critical configuration at depinning
\cite{RossoKrauth2001a,TheseAlberto}. The origin of recent progresses
in both numerics and field theory are thus related.  Clearly one would
like to test this novel field theory by calculating more universal
measurable quantities and study its properties.

In this paper we further explore the field theory constructed in
\cite{ChauveLeDoussalWiese2000a,LeDoussalWieseChauve2002,LeDoussalWieseChauve2002a}. We
study displacement correlations of more than two points. We find that
these correlations are {\it static}. Although physically natural, if
one wants quasi-static depinning to make sense, this manifests itself
through rather non-trivial massive cancellations in the time
dependence of multi-point diagrams. We elucidate these cancellations
and obtain as a consequence for a large class of diagrams much simpler
diagrammatic rules than previously anticipated. Basically, all time
integrals become almost trivial, resulting in a theory with ``quasi
static'' diagrams. We then apply these properties to the calculation
of universal observables.  One natural universal quantity is the
so-called width distribution of the interface.  Interestingly, to the
two lowest leading orders in $\epsilon=4-d$, the distribution coincides
with the one for a Gaussian theory with the full non-trivial
propagator $G(q)=1/q^{d+2 \zeta}$, $\zeta$ being the depinning
exponent. This is also the subject of a related publication
\cite{RossoKrauthLeDoussalVannimenusWiese2002}, where the distribution
is also measured numerically.  Here we give a detailed presentation
and also compute the higher connected cumulants of the displacement
field, i.e.\ deviations from the Gaussian. Some of these results are
quoted in \cite{RossoKrauthLeDoussalVannimenusWiese2002}.

In a second part we study the theory at the upper critical
dimension. The motivation is that no exact result is available to
confirm that $d_{\mathrm{uc}}=4$ (the only exactly solved limit
corresponding to fully connected models
\cite{DSFisher1985,Fisher1985b,VannimenusDerrida2001}).  Thus the
question of what is the upper critical dimension $d_{\mathrm{uc}}$ is
still debated, even though the field theory of depinning
\cite{NattermannStepanowTangLeschhorn1992,NarayanDSFisher1992b,%
NarayanDSFisher1993a,ChauveLeDoussalWiese2000a,LeDoussalWieseChauve2002,%
LeDoussalWieseChauve2002a,LeDoussalWiese2001} clearly predicts
$d_{\mathrm{uc}}=4$. Also, in the other class of depinning
transitions, the so called anisotropic depinning class with KPZ
nonlinearities, there is not even a convincing prediction for
$d_{\mathrm{uc}}$
\cite{TangKardarDhar1995,AlbertBarabasiCarleDougherty1998,LeDoussalWiese2002a}
and recent numerical studies have reopened the debate
\cite{RossoHartmannKrauth2002}.  Recently it has become possible to
study numerically depinning and statics in high dimensional spaces for
reasonable system sizes with better precision, allowing for the hope
to settle the issue of the upper critical dimension in the near future
\cite{RossoKrauth2001a,RossoKrauth2001b,RossoKrauth2002,RotersLubeckUsadel2002}.
It is thus important to give precise predictions for the behavior
predicted by the FRG, in order to compare with numerics.

Finally, we also clarify the issue of finite size scaling. In a
previous work, used in several simulations, the value $\omega
=-\epsilon /3$ was used for the finite size scaling exponent
\cite{RamanathanFisher1998,SchwarzFisher2002}. We find that the
correct value is instead $\omega =-\epsilon$. This may prove useful in
numerical studies \footnote{This has been tested for the data of
\cite{SchwarzFisher2002}. Even though there are no conclusive results,
$\omega =-\epsilon$ is favored over $\omega =-\epsilon /3$. We thank J.\ M.\
Schwarz for this private communication.}.

The paper is organized as follows. In Section \ref{sec2} we define the
model, briefly review the FRG method and field theory and define the
main observable of interest here, the width distribution. In Section
\ref{sec3} we compute the Laplace transform of the width distribution
in perturbation theory and find that to lowest order in $\epsilon$ it
coincides with a Gaussian Approximation.  This approximation is
introduced and further studied. Some results on Laplace Inversion are
given. In Section \ref{sec4} we go beyond the Gaussian Approximation
and compute higher connected cumulants. The detailed calculation of
the fourth cumulant (4-point connected correlation function of the
displacement field) is given, and the cancellations that occur in the
field theory are  studied. In Section \ref{sec5} we discuss the
upper critical dimension and in Section \ref{sec6} the finite size
scaling.  The effect of various boundary conditions is studied in
Appendix \ref{differentbound}.

\section{Model and observables}
\label{sec2}

\subsection{Model}

We study the over-damped dynamics
described by the equation of motion
\begin{eqnarray}\label{eqn.motion}
\eta \partial_t u_{xt} = c \nabla_x^2 u_{xt}  + F(x, u_{xt} ) + f
\label{eqmo1}
\end{eqnarray}
with friction $\eta$. Long range elasticity relevant for solid
friction at the upper critical dimension $d=2$, can be studied
replacing $c q^2 \to c |q|$. In presence of an applied force $f$ the
center of mass velocity is $v = L^{-d} \int_x \partial_t u_{xt}$. The
pinning force is $F(u,x) = - \partial_u V(u,x)$ and thus the second
cumulant of the force is
\begin{equation}\label{lf17}
 \overline{F(x,u) F(x',u')} = \Delta(u-u') \delta^d(x-x') \ ,
\end{equation}
such that $\Delta(u) = - R''(u)$ in the bare model, where $R(u)$ is
the correlator of the random potential. Random bond disorder is
modeled by a short range function $R(u)$, random field (RF) disorder
of amplitude $\sigma$ by $R(u) \sim - \sigma |u|$ at large $u$ and CDW
disorder by a periodic function $R(u)$.

\subsection{Review of FRG and field theory}
\label{s:review}
Let us briefly review the field theoretic approach, more details can
be found in \cite{LeDoussalWieseChauve2002}. The dynamical action
(MSR) averaged over disorder is given by $\rme^{-{\cal S}}$ with
\begin{equation}\label{3}
{\cal S}[u, \hat u] = \int_{xt} \hat u_{xt} (\partial_t - \nabla_x^2) u_{xt}
- \frac12 \int_{xtt'} \hat u_{xt} \Delta(u_{xt}-u_{xt'}) \hat u_{xt'}
\end{equation}
Here and below we denote $\int_x := \int \rmd^d x$, in Fourier $\int_k
:= \int \frac{\rmd^d k}{(2 \pi)^d}$ and $\int_t = \int \rmd t$.  The
FRG shows that the full function $\Delta(u)$ becomes relevant below
$d=d_{\mathrm{uc}}=4$ and a flow equation for its scale dependence has
been derived to one and two loops, in an expansion in $d=4 -
\epsilon$. In Ref.~\cite{LeDoussalWieseChauve2002} this was derived by
adding a small mass $m$ as an infrared cutoff and computing the flow
of disorder, defined from the effective action $\Gamma[u,\hat u]$ of
the theory, as $m$ decreases towards zero.  As in
\cite{LeDoussalWieseChauve2002} we will denote by $\Delta_0(u)$ the
bare disorder correlator, i.e.\ the one appearing in the action $\cal S$ in
(\ref{3}), and by $\Delta(u)$ the renormalized one, appearing in
$\Gamma[u,\hat u]$ which has a similar expression as (\ref{3}). The 
rescaled disorder is then defined by
\begin{equation} \label{resc}
\Delta(u) = \frac{1}{\epsilon \tilde{I_1}} m^{\epsilon - 2 \zeta}
\tilde{\Delta}(u m^{\zeta})\ ,
\end{equation}
where $I_1 = m^{- \epsilon} \tilde{I_1} = \int_k (k^2+m^2)^{-2}$
is the 1-loop integral. It was then shown in
\cite{ChauveLeDoussalWiese2000a,LeDoussalWieseChauve2002} that
(\ref{3}) leads to a  functional renormalization group equation
\begin{eqnarray}
-m \partial_m {\tilde \Delta}(u) &=&  (\epsilon - 2 \zeta) \tilde \Delta(u)
+ \zeta u \tilde \Delta'(u) \nn\\
&& - \frac{1}{2} \left[(\tilde \Delta(u) - \tilde
\Delta(0))^2\right]'' \nonumber \\&&
+ \frac{1}{2} \left[ (\tilde \Delta(u) - \tilde \Delta(0)) \tilde
\Delta'(u)^2 \right]'' \nn\\ 
&& + \frac{1}{2} \tilde \Delta'(0^+)^2 \tilde \Delta''(u)
\label{rgdisorder}\ .
\end{eqnarray}
up to $O(\Delta^4)$ terms. This equation implies that there are only
two main universality classes at depinning, a single RF fixed point
for interfaces and a periodic one for CDW type disorder
\cite{ChauveLeDoussalWiese2000a,LeDoussalWieseChauve2002}. Both
$\zeta$ and the fixed point function $\tilde \Delta^*(u)$ were
determined to order $O(\epsilon^2)$ for these classes
\cite{ChauveLeDoussalWiese2000a,LeDoussalWieseChauve2002}.

The important feature of the field theory of depinning is that
$\Delta(u)$ has a cusp-like non-analyticity at $u=0$. As was shown
in \cite{ChauveLeDoussalWiese2000a,LeDoussalWieseChauve2002}
calculations in the non-analytic theory (e.g.\ yielding
(\ref{rgdisorder})) are meaningfully performed using the
expansion:
\begin{eqnarray}
\Delta(u)=  \Delta(0) + \Delta(0^+) |u| + \frac{1}{2} \Delta(0^+)
u^2 + \dots . \label{exp}
\end{eqnarray}
Performing Wick averages yields the usual diagrams, except that their
actual values involve averages of e.g.\ sign functions of the
fields. Replacing everywhere $\sgn(u_t - u_{t'}) \to \sgn(t-t')$ is
justified for single component quasi-static depinning (i.e.\ in the
limit of vanishing velocity $v=0^+$). This yields diagrams with
sometimes complicated internal time and momentum dependences. We find
however that in some cases massive cancellations occur despite the
complications due to the time dependence between various diagrams,
contributing to the same observable.

\subsection{Universal distributions and observables}

To motivate the present study let us consider one specific example of
a universal observable, the width distribution of the configuration
at depinning (the so-called critical configuration). The width of a
configuration is defined {\it in a given disorder realization} as
\begin{equation}\label{1}
w^2 := \frac{1}{L^d} \int_x ( u_x - \overline{u})^2\ ,
\end{equation}
where $\overline{u} = \frac{1}{L^d} \int_x u_x$ is the center of mass
and $L^d$ the volume of the system. The basic observation is that the
sample to sample probability distribution $P(w^2)$ of $w^2$ is
expected to be universal, with a single scale set by the disorder
averaged second cumulant $\overline{w^2}$, i.e:
\begin{eqnarray}
P(w^2) =\frac{1}{ \,\overline{w^2}\,}
\,f\!\left(\frac{w^{2}}{\,\overline{w^2}\,}\right)\ .
\end{eqnarray}
$f(z)$ is a universal function. This holds for thermal averages in a
number of finite temperature problems of pure systems
\cite{PlischkeRaczZia1994,FoltinOerdingRaczWorkmanZia1994}. Here we
show that it also holds for depinning at $T=0$ and compute the
distribution, first within a simple Gaussian approximation and second
within the $\epsilon$-expansion. In the process we study higher point
correlation functions in the field theory of depinning, define
specific universal ratios of these, describing deviations from
Gaussian behavior and compute them.

Before turning to actual calculations let us first summarize the
general spirit of the method and discuss the question of the
universality of such observables. The hallmark of a renormalizable
theory is that if one expresses the correlation functions in an
expansion in the {\it renormalized} disorder $\Delta$, then the
resulting expressions are UV finite, equivalently they have a well
defined continuum limit, independent of short scale details. On a
technical level, this can be achieved by computing correlations in
standard perturbation theory to a given order in powers of $\Delta_0$,
and then using the relation between renormalized disorder $\Delta$ and
bare one $\Delta_0$ to the same order, or equivalently through the
definition of appropriate counter-terms \footnote{A subtle point 
in that construction is that if one defines $\Delta_0$ perturbatively
from $\Delta$ to a given order then $\Delta_0$ is not the original
bare action (which is analytic) thus there is no contradiction in
$\Delta_0$ being non-analytic. In a sense introducing $\Delta_0$ is
just a trick to express a closed equation for the flow of $\Delta$ to
the same order.}. Here, we restrict ourselves to calculations at
dominant order in $\epsilon$ and thus using either $\Delta$ or
$\Delta_0$ makes no difference. Beyond the Larkin scale, however,
these are non-analytic functions, which is crucial.

In the limit of large scales or large system sizes, the fixed point
form reached by the rescaled $\Delta$ implies that the resulting
observable, e.g.\ the width distribution, is {\em
universal}. Universal means that these quantities do not depend on the
short scale details. However they {\it do depend} on the details of
the large scale infrared (IR) cutoff, i.e.\ of the type of chosen
boundary conditions. Here we focus on {\it periodic boundary
conditions} \cite{caveat}, also of interest for numerical simulations
\cite{RossoKrauthLeDoussalVannimenusWiese2002}, although we sometimes
give results for other types, for instance for the massive IR cutoff
described in the previous paragraph.

Since the FRG method developed in
\cite{ChauveLeDoussalWiese2000a,LeDoussalWieseChauve2002} and
summarized above uses a mass as IR-cutoff and defines disorder
vertices at zero momentum, one should be careful in calculations with
e.g.\ periodic boundary conditions. Since we only compute observables
either to dominant order in $\epsilon$, or within a 1-loop
approximation, it is easy to make the necessary replacements, as will
be indicated below. For instance the 1-loop FRG equation remains
identical to the two first lines of (\ref{rgdisorder}), the only
changes being that \begin{enumerate}
\item $-m \partial_m \tilde{\Delta}$ has to be replaced by $L
\partial_L \tilde{\Delta}$.
\item  $m \to
1/L$ in the definition of the rescaled disorder.
\item the 1-loop integral $I_{1}=\int_{k}\frac{1}{(k^{2}+m^{2})^{2}}$
entering into the definition of the rescaled disorder has to be
replaced by its homologue for periodic boundary conditions
\footnote{To extend the 2-loop FRG (\ref{rgdisorder}) to that periodic
b.c.~scheme, one would have to reevaluate 2-loop integrals with
discrete sums, or carry other changes in the scheme. This however goes
beyond the present 1-loop study.}:
\begin{equation}
I_1 := \sum_{k}\frac{1}{(k^{2})^{2}} \equiv L^{-d} \sum_{n \in \mathbb{Z}^d, n
\neq 0} \frac{1}{(2 \pi n/L)^4}
\end{equation}
used  below.
\end{enumerate}

\section{Width distribution: Perturbation theory and Gaussian
Approximation} \label{sec3}

Let us start by giving the simplest approximation for
this distribution. It can be derived in two ways:
(i) perturbation theory in the renormalized theory to
lowest order in $\epsilon$ (ii) a simple Gaussian
approximation. In the end this will motivate
going further, i.e.\ studying deviations from the
Gaussian approximation.

\subsection{Perturbation theory} We now study perturbation theory.  To
compute the width distribution using the dynamical field theoretic
method \cite{ChauveLeDoussalWiese2000a,LeDoussalWieseChauve2002} one
can start from the Laplace transform
\begin{equation} \label{2}
W(\lambda) = \overline{\rme^{-\lambda w^2}}
\end{equation}
with $w^2 =  \sum_x (u_x -\overline {u})^2$. Here and below we
omit the global multiplicative factor $L^{-d}$ in the definition
of $w^2$, since in the end we will always normalize the
distribution $P(w^2)$ by fixing its first moment to unity (in
(\ref{2}) it can be absorbed by a rescaling of $\lambda$).

\begin{figure}[t]  \fig{\columnwidth}{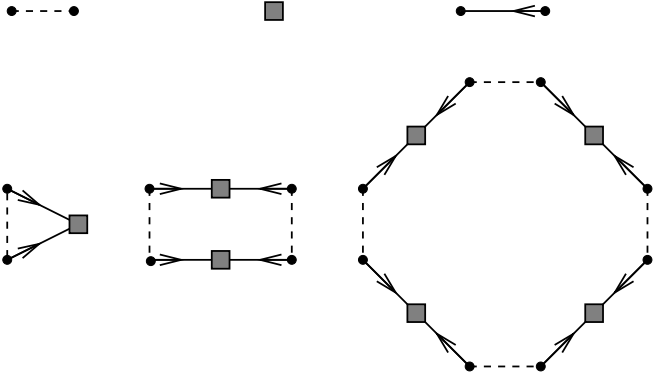}
\caption{Examples of contributions to (\protect\ref{2}), terms
$\overline{w^2}^c$, $\overline{(w^2)^2}^c$ and $\overline{(w^2)^4}^c$
(bottom), together with the vertices for disorder (top left), $w^2$
(top center), and response function (top right).}\label{fig1}
\end{figure}

Expanding in powers of the correlator of the pinning force $\Delta(u)$
(to lowest order this is equivalent to $\Delta_0$ see above), one
finds that to leading order $\ln W(\lambda)$ is the sum of all
connected 1-loop diagrams, as represented in Fig.~\ref{fig1}.
The loop with $n$ disorder vertices and $n$ insertions of $w^2$ is
\begin{equation}
\frac1{2 n} \sum_q \left(\frac{- 2 \lambda \Delta(0)}{(q^2)^2} \right)^n
\label{one}
\end{equation}
Here and below the sums over $q$ thus runs over a $d$-dimensional
hypercubic lattice with spacing $\frac{2\pi}{L}$, and the 0-mode
is excluded, as appropriate for periodic boundary conditions. If
one were now to resum (\ref{one}) over $n$ one would find:
\begin{equation}\label{il3}
W(\lambda) = \prod_q \left( 1 + 2 \lambda G(q) \right)^{-1/2}
\label{final}
\end{equation}
with $G(q) = G_{\mathrm{Larkin}}(q)= \Delta(0)/q^4$, a Larkin model type
result if interpreted as naive perturbation theory (i.e.\ if
$\Delta(0)$ was interpreted as the bare original disorder rather
than the renormalized one). The correct procedure implies that
$\Delta(0)$ is the running renormalized disorder $\Delta(0) \to
\Delta_l(0) = (\epsilon \tilde{I}_1)^{-1} \rme^{(2 \zeta -
\epsilon) l} \tilde{\Delta}^*(0)$ where $\tilde{\Delta}^*(0)$ is
the (non-universal) value of the fixed point
\cite{ChauveLeDoussalWiese2000a,LeDoussalWieseChauve2002}.
Although $l=\ln(L)$ for the zero momentum disorder vertex, one
notes that a momentum $q$ flows in each vertex and one should take
care of this by setting  $l \to \ln(1/q)$. This yields finally (\ref{il3}) with
\begin{equation}\label{il3p}
G(q) = C/q^{d + 2 \zeta}
\end{equation}
where the value of $C$ is non-universal and fixed by
$\overline{w^2}$. As will become clear in the following section, the
appropriate choice for $G(q)$ is the 2-point finite-size
scaling-function $G_L(q) = C/q^{d+2 \zeta} g(q L)$ with $g(0)=1$.  The
difference between the two above-mentioned choices for $l$ simply
amounts to the two different limits of small, or large $q L$. However
to lowest order in $\epsilon=4-d$ they are identical. (For a
calculation of the scaling function to next order in $\epsilon$ see
Appendix J in Ref. \cite{LeDoussalWieseChauve2002}.)

\subsection{Gaussian Approximation and beyond}\label{beyondGauss}
A more general approach consistent with the previous calculation is
the following. We first note that the above result (\ref{final}) would
be exact if the distribution of the displacement fields $u$ were {\em
Gaussian}. It can thus be called the Gaussian Approximation (GA). To
understand why it was obtained here let us consider simply the second
connected cumulant of the WD $ \overline{(w^2)^2}^c = \overline{
(w^2)^2 } - \overline{(w^2)}^2$.  This cumulant however is not
connected w.r.t. the $u$, and thus there is an exact relation:
\begin{equation}\label{conn}
\overline{( w^2)^2}^c = \int_{xy} \left(  2 G_{xy}^2 + \overline{ u_x^2
u_y^2 }^c \right) 
\end{equation}
where here $G_{xy} = \overline{u_x u_y}$ is the exact disorder
averaged 2-point function.  The first term is just Wick's
theorem and would be the full result if the measure of the $u$
were Gaussian. Analogous formulae exist for higher cumulants: the
first term on the r.h.s.\ of (\ref{conn}) generalizes into
\begin{equation}\label{conngen}
\overline{(w^2)^n}^c\big|_{\mathrm{GA}} = 2^{n-1} (n-1)! \int_{x_1,..x_n}
G_{x_1 x_2} G_{x_2 x_3}\dots G_{x_{n} x_1}
\end{equation}
as a simple consequence of Wick's theorem.  This is again easily
resummed into (\ref{final}) which would thus be exact if the measure
of $u$ were exactly Gaussian. Note that all results of
\cite{PlischkeRaczZia1994,FoltinOerdingRaczWorkmanZia1994} for pure
Gaussian theories can also easily be obtained by the present
resummation method (temperature replacing disorder).  Thus in the GA,
the $G(q)$ appearing in (\ref{final}) is the exact 2-point
function. It can be tested in a simulation by inserting the measured
2-point function in (\ref{final}). In the large but finite system
size limit it takes the scaling function form $G_L(q)$ discussed
above.

When comparing to the numerical results for the width distribution, it
turns out that the GA is a surprisingly good approximation even down
to $d=1$. This is discussed in details in
\cite{RossoKrauthLeDoussalVannimenusWiese2002}. However, we do not
expect the GA to be exact. It is thus interesting to compute the
deviation $D= \int_{xy} \overline{ u_x^2 u_y^2 }^c$ for the second
cumulant of $w^2$ in (\ref{conn}). It is computed below and found to
be of order $\epsilon^4$ while the GA contribution (first term in
(\ref{conn})) is of order $\epsilon^2$ (since $G \sim
\epsilon$). Similarly the GA contribution to (\ref{conn}) is
$O(\epsilon^n)$, while the deviations are found to be $O(\epsilon^{2
n})$. This can be summarized as $u= \sqrt{\epsilon} u_0 + \epsilon
u_1$ where $u_0$ is a Gaussian random variable of $O(1)$ and $u_1$ a
non-Gaussian one of $O(1)$.

Computing deviations from the GA is thus one motivation to compute
higher point correlations.

\subsection{Laplace inversion}

Before doing so, let us discuss how the distribution $P(w^{2})$ is
obtained through an inverse Laplace-transform, as
\begin{equation}\label{il}
P (w^{2}) = \oint \frac{\rmd \lambda}{2\pi i}\, W (\lambda)\,
\rme^{\lambda w^{2}}
\end{equation}
Noting that in $d=1$ (\ref{final}) can also be written as
\begin{equation}\label{final2}
W(\lambda) = \prod_{q> 0} \left( 1 + 2 \lambda G(q) \right)^{-1}\ ,
\end{equation}
this is equivalent to 
\begin{equation}\label{finalLapTran}
P (w^{2}) = \sum_{p>0}\rme^{-\frac{w^{2}}{2 G (p)}} \frac{1}{2 G
(p)}\prod_{q>0, q\neq
p} \left(1-\frac{G (q)}{G (p)} \right)^{-1}\ .
\end{equation}
This formula shows that for large $w^{2}$, the distribution is
dominated by the first term $p=1$, and in practice summing the first
few terms gives an excellent approximation.

It is instructive to apply (\ref{finalLapTran}) to a random walk,
where $G (q)=1/q^{2}$. Using that ($n\in \mathbb{N}$)
\begin{equation}\label{help1}
\prod_{n>0} \left(1-\frac{x^{2}}{n^{2}} \right) = \frac{\sin (\pi
x)}{\pi x}
\end{equation}
one finds in terms of the width $\overline{w^{2}}$
\begin{equation}\label{rw}
P (w^{2}) = \overline{w^{2}}\, \frac{\pi^{2}}{3} \sum_{n>0} n^{2}
(-1)^{n+1} \rme^{-\frac{\pi^{2}}{6}\frac{w^{2}}{\,\overline{w^{2}}\,}n^{2}}\ .
\end{equation}

For $d>1$ the situation is more complicated. Writing
\begin{equation}\label{il2}
P (w^{2}) = \oint \frac{\rmd \lambda}{2\pi i} \,
\rme^{w^{2}\lambda} \prod_{q,q_x>0} (1+2\lambda G (q))^{-1}\ .
\end{equation} 
we have e.g.\ at least multiplicity $2d$ for each factor in
(\ref{il3}), as long as no component vanishes, but this multiplicity
may even be higher, as can be seen from the following solutions of the
diophantic equation (for 2 dimensions) $1^{2}+7^{2}=5^{2}+5^{2}$,
$6^{2}+7^{2}=9^{2}+2^{2}$, and many more. Let us define the class
${\cal C} (q)$ as
\begin{equation}\label{mul}
p \in {\cal C} (q) \quad \mbox{if} \quad  p^{2} = q^{2}\ .
\end{equation}
Let us index these classes by $\alpha$, and introduce an order
\begin{equation}\label{lf18}
{\cal C}_{\alpha} < {\cal C}_{\alpha '} \quad \mbox{if}\ q\in {\cal
C}_{\alpha} \mbox{ and } q' \in {\cal C}_{\alpha '} \quad \Rightarrow
|q|<|q'|
\end{equation}
The number of elements of each class is defined as
\begin{equation}\label{lf19}
| {\cal C}_{\alpha}| := \mbox{number of elements in }{\cal C}_{\alpha }
\end{equation}
We further define
\begin{equation}\label{lf1}
q_\alpha := \mbox{any element out of } {\cal C}_{\alpha}\ .
\end{equation}
Note that since for $p\in{\cal C}_{\alpha}$ and $p\neq 0$ (by
definition we exclude $p=0$)  $| {\cal C}_{\alpha}|$ is always even.
(\ref{il2}) can then be rewritten as 
\begin{equation}\label{il4}
P (w^{2}) = \oint \frac{\rmd \lambda}{2\pi i} \, \rme^{w^{2}\lambda}
\prod_{\alpha } (1+2\lambda G (q_\alpha ))^{-|{\cal C}_{\alpha}|/2}\ .
\end{equation}
There are poles at $\lambda =-[2 G (q_{\alpha})]^{-1}$. The sum over these
poles can  after partial integration be written as \begin{widetext}
\begin{equation}\label{lf20}
P (w^{2}) = \sum_{\alpha} \frac{1}{(|{\cal C}_{\alpha}|/2-1)!}
\left(\frac{1}{2G (q_{\alpha})} \right)^{|{\cal C}_{\alpha}|/2} \left(
\frac{\partial}{\partial \lambda }\right)^{|{\cal C}_{\alpha}|/2-1}
\left.\left[ \rme^{w^{2}\lambda}\prod_{\alpha '\neq \alpha} (1+2
\lambda G (q_{\alpha}))^{-|{\cal C}_{\alpha}|/2}
\right]\right|_{\lambda = - [ 2G (q_{\alpha })]^{-1}}
\end{equation}
\end{widetext}

\section{Higher point correlations in depinning field theory}
\label{sec4}

In this section we analyze how one can compute higher correlations in
the depinning field theory and obtain simple diagrammatic rules for
doing so. These are illustrated on the 4-point function. Specific
calculations and results will be given in the next Section.

\subsection{Preliminaries}
We want to compute at $T=0$, using the dynamical action ${\cal S}$ in
(\ref{3}) , e.g.\ the 4-point expectation value, connected
w.r.t. disorder (and $u$) as defined in the previous sections:
\begin{equation}\label{lf2}
 \int_{xy} \overline{u_{xt}^2 u_{yt}^2}^c = \int_{xy} \left< u_{xt}^2
u_{yt}^2 \right>_c \ .
\end{equation}
Similar formulas hold for higher correlation functions. This is
identical to a connected expectation value w.r.t. the action ${\cal
S}$, denoted hereafter $\left<\dots\right>_c$.

The first step is to show that correlations can all be expressed
as:
\begin{eqnarray}\label{lf21}
 \left< u_{xt}^2 u_{yt}^2 \right>_c &=&
\int_{x_i, t_i < t} {\cal R}_{xt,x_1 t_1} {\cal R}_{xt,x_2 t_2}
{\cal R}_{yt,x_3 t_3} {\cal R}_{yt,x_4 t_4} \nonumber \\
&&\qquad \times
\Gamma^{(4)}_{\hat u \hat u \hat u \hat u}(x_1 t_1, x_2 t_2, x_3 t_3,
x_4 t_4) \qquad \qquad
\end{eqnarray}
Here ${\cal R}$ is the exact response function and $\Gamma$ the exact
effective action (sum of 1PI graphs) (with the choice $\rme^{-\Gamma}$
for the probability and $\Gamma^{(4)}$ is symmetric). This is the
standard relation between connected correlation functions and the
effective action (i.e.\ 1-particle irreducible vertex functions
IVF). The simplification here is that a priori the exact 2-point
correlation function and vertices such as $\Gamma^{(4)}_{u u \hat u
\hat u}$ could also contribute, but their contribution vanishes for
$T=0$ at the depinning threshold. This is because $\left< u_{x t}
u_{x,t_1}\right>$ is time independent there, and then statistical tilt
symmetry implies that all IVF's with at least one external $u$ leg
carrying frequency $\omega$ vanish when this frequency is set to zero
(see section II A in \cite{LeDoussalWieseChauve2002}). The above
formula (\ref{lf21}) generalizes straightforwardly to any connected
$2n$-point correlation function of the field $u$ in terms of
$\Gamma^{(2n)}_{\hat u \dots \hat u}$.

Next one can compute $\Gamma^{(4)}_{\hat u \hat u \hat u \hat u}$ in
perturbation, using the diagrammatic rules for the non-analytic action
arising from the expansion (\ref{exp}). Let us denote $E_{\hat u}$ the
number of $\hat u$ external legs, $n_v$ the number of (unsplitted)
vertices $\Delta_0$, $n_I$ the number of internal lines (response
functions) and $L$ the number of (momentum) loops. Then one has $2 n_v
- n_I = E_{\hat u}$ and $L= 1 + n_I - n_v$. Here one has $E_{\hat
u}=4$ and thus the lowest order contribution has $n_v=4$, $n_I=4$,
i.e.\ it is the 1-loop square (since two vertices $n_v=2$ implies
$n_I=0$ and is disconnected, three vertices $n_v=3$ implies $n_I=2$
and is 1-particle reducible). The two loop corrections are diagrams
with 5 vertices and so on. Similarly $\Gamma^{(2n)}_{\hat u \dots
\hat u}$ to lowest order is the 1-loop $2n$ polygon diagram.

\begin{figure}[b]
\Fig{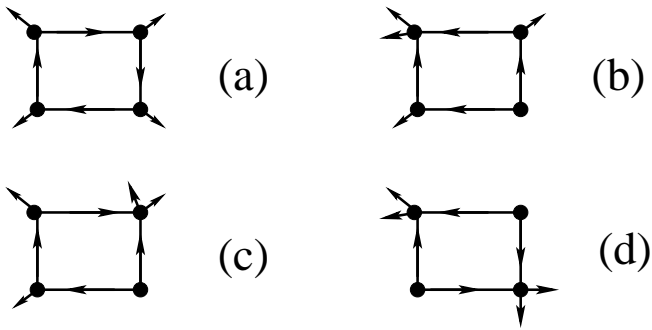}
\caption{The four 1-loop diagrams with unsplitted vertices which contribute to
the 4-point irreducible vertex function $\Gamma^{(4)}_{\hat u \hat u \hat u \hat u}$}
\label{fig2}
\end{figure}

Since two $\hat u$ fields must come out (arrows) of each vertex, there
are four possible diagrams corresponding to the 1-loop square, shown
on figure \ref{fig2}. Each line entering a vertex corresponds to one
derivative of the $\Delta(u)$ function of the vertex. Thus from
(\ref{exp}) we see that (a) is proportional to $\Delta'(0^+)^4$, (b)
and (c) to $\Delta''(0^+) \Delta'(0^+)^2 \Delta(0)$ and (d) to
$\Delta''(0^+)^2 \Delta(0)^2$. However, as we will show below using
the so-called mounting property only (a) is non-zero.

We found two ways to compute diagram (a) (as well as any other similar
diagrams), a systematic but complicated way, and a simple way which
uses a very important property and not yet fully elucidated of the
field theory of depinning, the ``quasistatic property'' described
below. To appreciate the extent of the cancellations involved in this
drastic simplification, we start by sketching the systematic method.

\begin{figure}
\centerline{\fig{1\columnwidth}{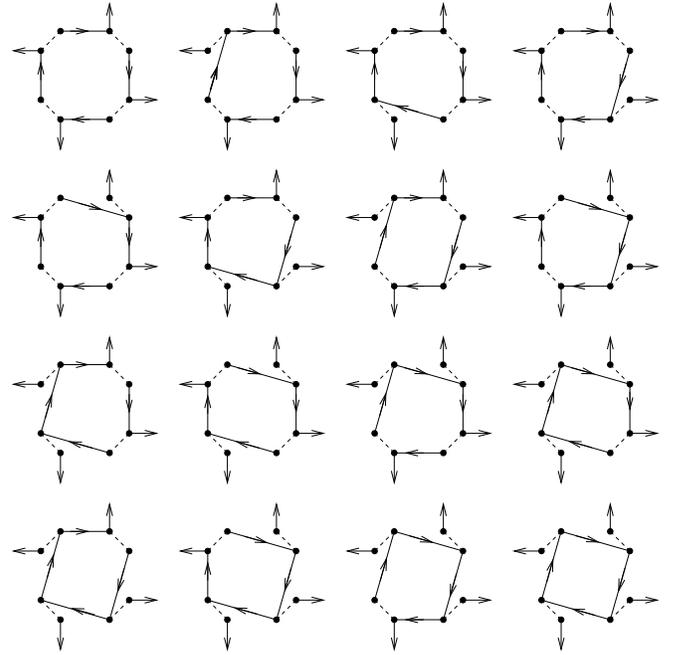}}
\caption{The 16 1-loop diagrams with splitted vertices which correspond
to diagram (a) in Fig. \ref{fig2}. The last one, which contains
an acausal loop and thus vanishes, is added here 
for future convenience.}
\label{ato16}
\end{figure}

To perform an actual calculation, since each $\Delta$ vertex involves 
fields at two times, one must switch to the splitted diagrammatics, as
described in \cite{LeDoussalWieseChauve2002}.  Diagram (a) then
becomes the sum of 16 splitted diagrams (two choices per vertex)
represented in Fig.~\ref{ato16} . Note that the last one is zero since it
involves an acausal loop. That leaves 15 non-zero and non-trivial
diagrams.

These diagrams correspond to the following. One first expands
 ${\cal S}^4/4!$ using (\ref{exp}), which gives, schematically:
\begin{equation}
\frac{\Delta'(0^+)^4}{2^4 4!} \hat u_1 \hat u_2 s_{12} u_{12} \hat
u_3 \hat u_4 s_{34} u_{34} \hat u_5 \hat u_6 s_{56} u_{56} \hat
u_7 \hat u_8 s_{78} u_{78}\ .
\end{equation}
In shorthand notations $\hat u_1=\hat u_{x_1,t_1}$, $\hat u_2=\hat
u_{x_1,t_2}$, $u_{12}=u_1-u_2$, $s_{12}=\sgn(t_1-t_2)$ omitting all
space and time integrals. On then carries the Wick
contractions,  yielding
\begin{eqnarray}
&& \frac{\Delta'(0^+)^4}{4!} 2 \hat u_1 \hat u_3 \hat u_5 \hat u_7
s_{12} s_{34} s_{56} s_{78} \nonumber \\
&& \quad \times (R_{32} - R_{42})(R_{54} - R_{64})(R_{76} - R_{86})(R_{18} -
R_{28})\ .\nonumber \\
\end{eqnarray}
$R_{ij}=\left< \hat{u}_{i}u_{j} \right>$ is the free response
function.  The factor of $2$ comes from the two possible time
orientations of the loop.  Expanding the product of response functions
yields the 16 diagrams represented in Fig.\ref{ato16}, where space and
time labels are ordered turning clockwise around the momentum
loop. For illustration let us indicate the full expression of the
first diagram in Fig.\ref{ato16}, in momentum space:
\begin{eqnarray}
&& \Gamma_{(\mathrm{a}1)}(p_{12},t_1;p_{23},t_3;p_{34},t_5;p_{41},t_7)= \\
&&   \Delta'(0^+)^4 \int_{t_2 t_4 t_6 t_8} \sgn(t_1-t_2)
\sgn(t_3-t_4) \sgn(t_5-t_6) \nonumber \\
&&\qquad \times  \sgn(t_7-t_8) R_{p_1,t_3-t_2}
R_{p_2,t_5-t_4} R_{p_3,t_7-t_6} R_{p_4,t_1-t_8}\nonumber \\
\end{eqnarray}
with $p_{ij}=p_i-p_j$ the entering momenta and
$R_{p,\tau}=\theta(\tau)e^{-p^2 \tau}$ the free response function.
Because of the sign functions the evaluation of these integrals,
and of the other 14 non-vanishing diagrams is very tedious and was
handled using Mathematica. Adding all diagrams, massive
cancellations occur. The final result is very simple and given
below.

Let us now explain the simple method and the properties of the
theory which lead to it.

\subsection{Theorems and other properties}

The simple way to compute the 4-point correlations (and higher ones)
at depinning is based on the very important following
conjectured property:

\medskip

\noindent \smallskip \underline{{\bf Quasistatic property 1
}:}\nopagebreak

All correlation functions $\langle u_{x_1 t_1} \dots u_{x_{2 n} t_{2
n}}\rangle $ computed using the diagrammatic rules of the quasistatic
field theory of depinning at zero temperature and exactly at
threshold, are {\bf independent of all time arguments}.

\medskip

Using relations such as (\ref{lf21}) for arbitrary times shows that an
equivalent way to state this property is the following:

\medskip

\noindent \smallskip \underline{{\bf Quasistatic property 2
}:}\nopagebreak

All $\Gamma^{(2n)}_{\hat u \dots \hat u}(x_1 t_1, \dots, x_{2n}
t_{2n})$ are {\bf independent} of $t_1,\dots ,t_{2n}$.

\medskip

This property, which appears as a physical requirement for the correct
field theory of depinning, implies non-trivial properties of the
diagrammatics. Although we will not attempt to prove it here in full
generality, we have checked it on many examples, and believe that it
works. We encourage the reader to contribute a valid proof. We will
however state and prove some easier and useful properties below.

Once the properties 1 and 2 are accepted, evaluation of the diagrams
drastically simplifies, thanks to the following trick: Since the
result does not depend on external times, one can take these times
mutually infinitely separated, with some fixed (and arbitrary)
ordering. Then one can integrate easily over all internal times since
the order at each vertex is then specified and each sign function has
a fixed value. One recalls that in the splitted diagrammatics all
non-vanishing $T=0$ diagrams are {\em trees} (see section II A in
\cite{LeDoussalWieseChauve2002}).  This can be seen on the fifteen
non-vanishing diagrams of Fig.\ref{ato16}. Thus integrating
independently along each tree starting from the leaves yields one
correlation function per link, since $\int_{t'} R_{q,t-t'} =
1/q^2$. Performing this calculation on all fifteen diagrams of
Fig.\ref{ato16} shows that they cancel pairwise since they differ only
by a global sign, with the exception of the acausal diagram which is
zero. Thus the final result is the same as if one had kept only $(-1)$
times the acausal graph!

Before giving the final result below, let us now state
the easier properties.

\medskip

\noindent \smallskip \underline{{\bf Theorem 1} (mounting
trick):}\nopagebreak

A diagram which contributes to an $u$-independent vertex function is 0
if it contains a vertex, into which no response-function enters.

Examples are diagrams (b), (c) and (d) on figure \ref{fig2}. This
theorem thus ensures that only (a) is non-vanishing.

\smallskip \leftline{\underline{\bf Proof:}} \nopagebreak The
following figure demonstrates the principle. Note that it may be part
of a larger diagram. Especially, there may be more response-functions
entering into the upper disorder. The statement is that
\begin{equation}\label{lf3}
\int \rmd t\ \left(\diagram{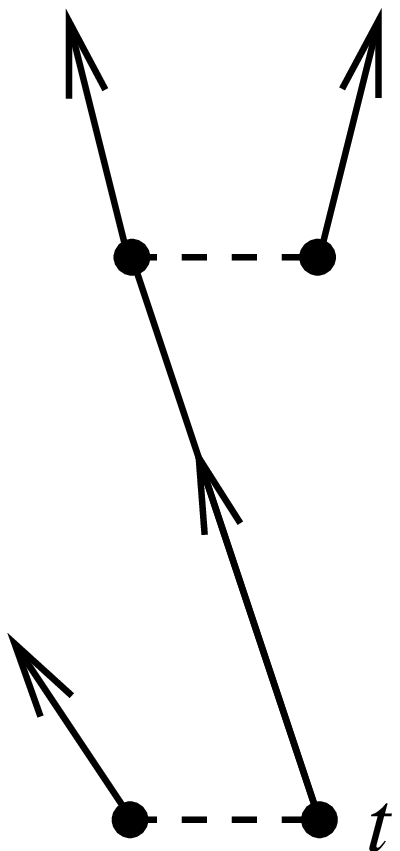} + \diagram{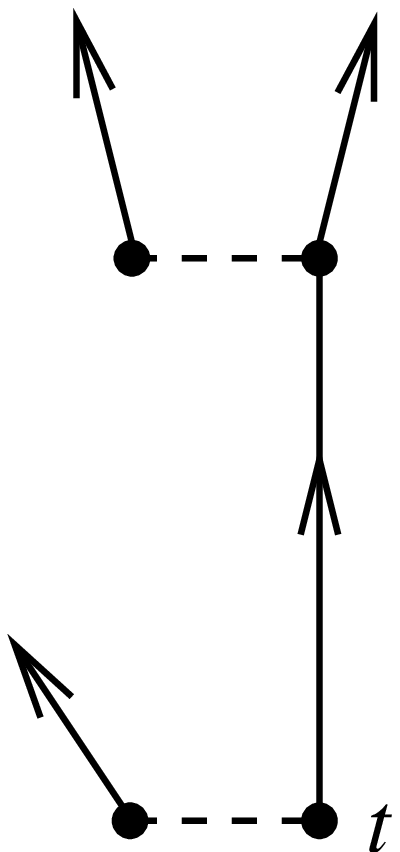}  \right)=0\ .
\end{equation}
Since no response-function enters into the lower disorder $\Delta
(u-u')$, due to the assumptions this gives $\Delta (0)$, with no
dependence on time. Thus
one can freely integrate over the response-function starting at time
$t$. This integral for both diagrams is $\int \rmd t\, R
(k,t)=1/k^{2}$. The difference in sign comes from deriving the two
different ends of the upper disorder vertex. Thus both contributions
exactly cancel.

Thus $\Gamma^{(4)}_{\hat u \hat u \hat u \hat u}(x_1 t_1, x_2 t_2, x_3 t_3, x_4 t_4)$
is given only by graph (a). We have not found a complete proof that
it is independent of external times, but we can prove the weaker

\smallskip \leftline{\underline{\bf Lemma 1:}} \nopagebreak
\begin{equation}\label{lf5}
\Gamma^{(4)}_{\hat u \hat u \hat u \hat u}(x_1 t_1, x_2 t_2, x_3 t_3, x_4 t_4)
\end{equation}
(see graph (a) on figure \ref{fig2}) is independent of the most
advanced time.

\smallskip \leftline{\underline{\bf Proof:}} \nopagebreak First
suppose that a response-function enters at the most-advanced time
$t'$. Then there is the following cancellation
\begin{equation}\label{lf6}
\int\rmd t\ \left(\diagram{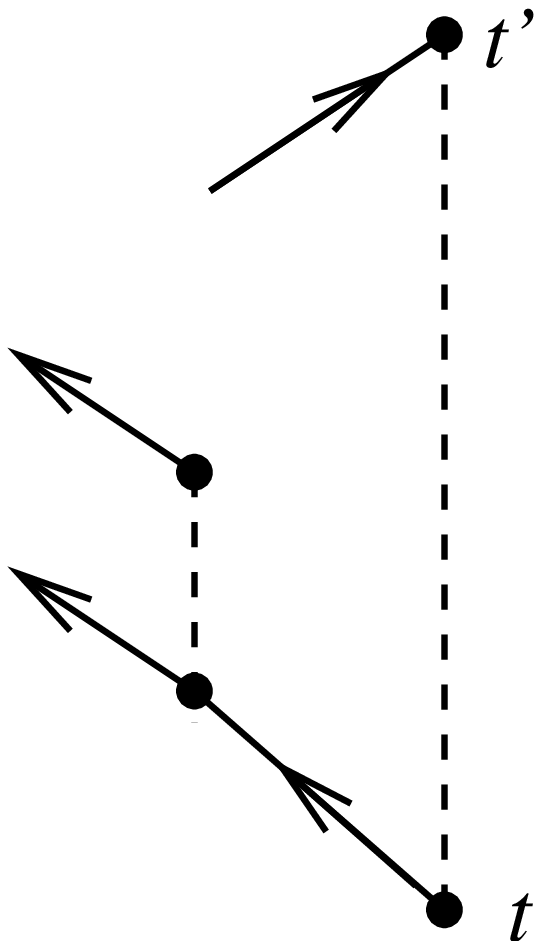}+\diagram{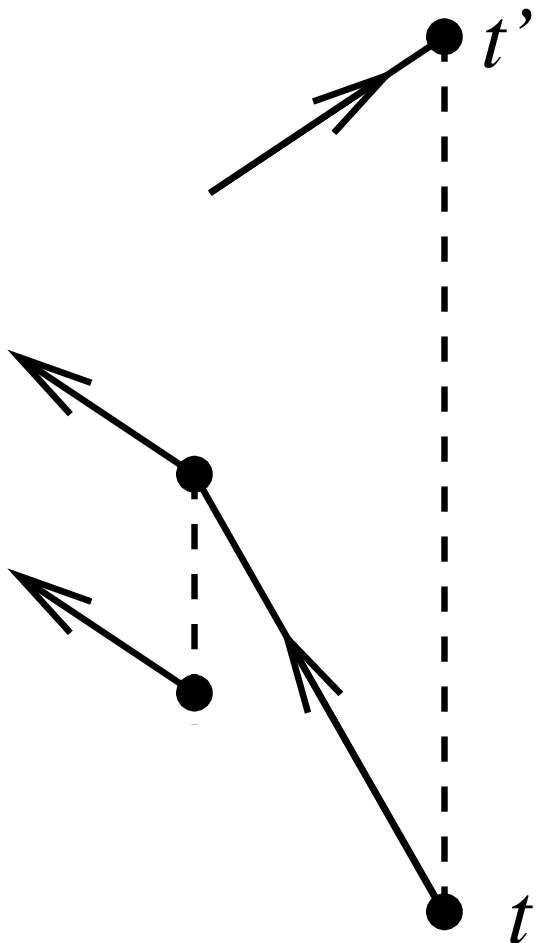} \right) =0
\end{equation}
The mechanism is the same as in the proof of theorem 1; since by
assumption $t'$ is the most advanced time, the
argument of the right-most disorder can never change sign, and  can be
integrated over, even though it is odd, i.e.\ $\sim \Delta' (0^{+})$.
Thus the remaining diagrams have the structure
\begin{equation}\label{lf7}
\diagram{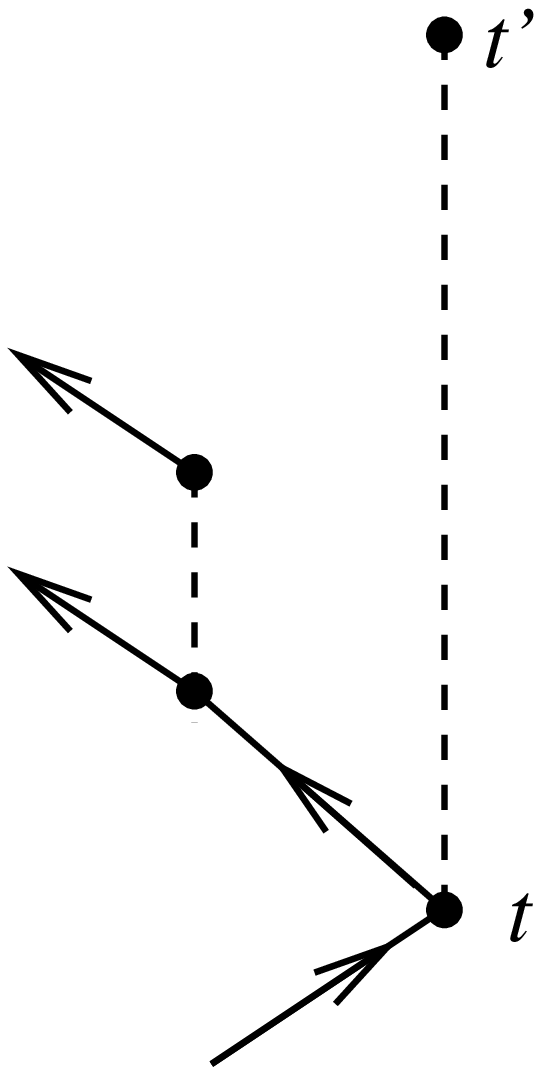}\ .
\end{equation}
This diagram  is independent of $t'$, as long as $t'$ is the
largest (external) time.

Note that for a loop made out of two disorders, there is only one diagram
remaining, namely
\begin{equation}\label{lf8}
\diagram{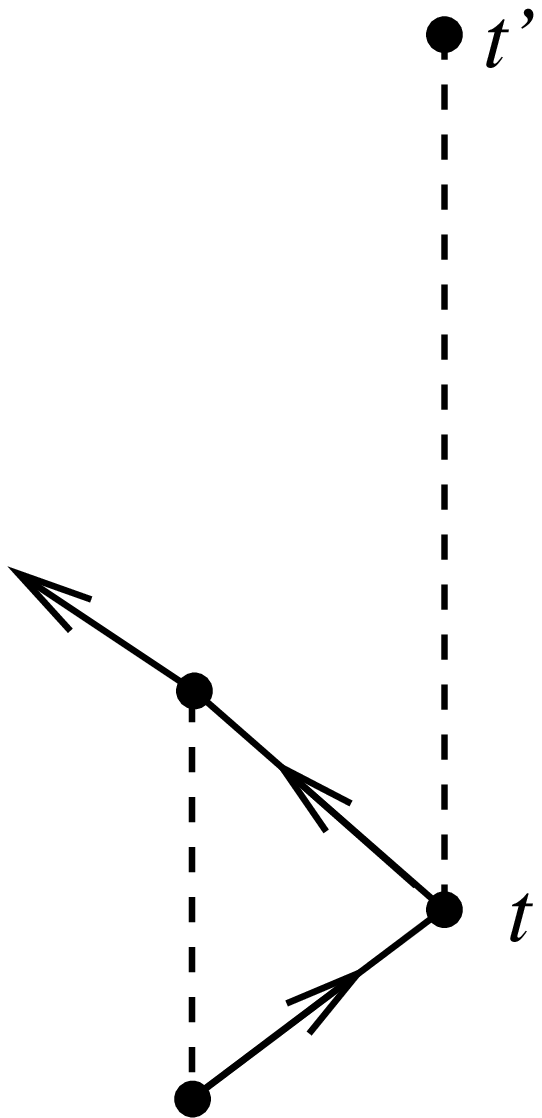}\ .
\end{equation}
It is manifestly time independent.

\smallskip \leftline{\underline{\bf Lemma 2:}}\nopagebreak
\begin{equation}\label{lf9}
\Gamma^{(4)}_{\hat u \hat u \hat u \hat u}(x_1 t_1, x_2 t_2, x_3 t_3, x_4 t_4)
\end{equation}
(see graph (a) on figure \ref{fig2}) is independent of the differences
in time, if those are very large.

\smallskip \leftline{\underline{\bf Proof:}}\nopagebreak By
inspection, one finds that by taking the external times, i.e.\ one
time at each disorder, (infinitely) far apart, the remaining integrals
become unambiguous. Thus integrating over the response-functions does
not leave any time-dependence.

As mentioned above, calculation of the diagram (a) becomes
then possible and one finds the

\smallskip \leftline{\underline{\bf Property} (missing acausal
loop):}\nopagebreak The diagram (a) is given by $(-1)$ times the
acausal loop, if there one replaces each response-function by a
correlation-function.

Intuitively this means, that if the acausal loop would give a
contribution, then all diagrams would cancel. This seems to be a
general mounting theorem.

\begin{figure}[t]
\Fig{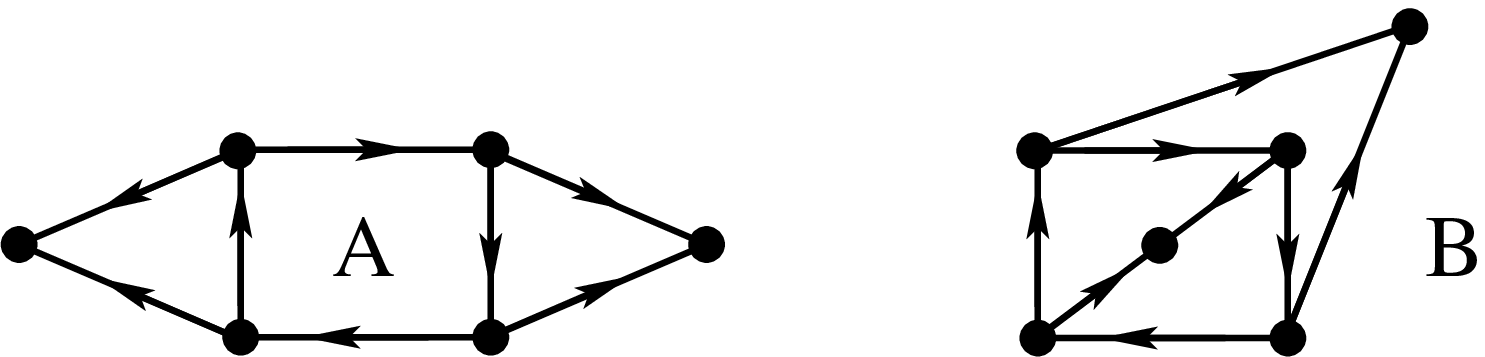} \caption{The two contributions to $\left< u^{4}
\right>_{c}$ at leading order.}  \label{fig3}
\end{figure}

\smallskip \leftline{\underline{\bf Check:}}\nopagebreak
One can calculate diagram
A on Fig.\ \ref{fig3} explicitly using none of the above theorems or
conjectures. The result is a formidable expression, which has to be
integrated over momenta. By evaluating it for given values of the
momenta (not even necessarily conserving momentum), one can compare
with the prediction of  property 5. We found both expressions to be
equal for any randomly chosen values of the momenta.

The properties described here suggest the following

\smallskip \leftline{\underline{\bf Property }(any loop):}\nopagebreak

All graphs can be computed to any number of loops, using
generalizations of the above rules.

This is not attempted here but preliminary investigation suggests that
the same mechanism holds for two loops with some simple end result
related to the signs of possible ''fermion loops''.

\section{Final result for the fourth cumulant and universal ratio}

In this Section we compute the fourth cumulant
\begin{eqnarray} \label{lf22}
D &=& \int_{xy} \left< u_{xt}^2 u_{yt}^2 \right>_c
\end{eqnarray}
as well as the ratio (kurtosis):
\begin{eqnarray}
R=\frac{D}{2 \int_{xy} G_{xy}^2 }
\end{eqnarray}
which, according to the discussion in Sections II and III
is expected to be universal and characterizes
the deviations from the Gaussian approximation
(for which $R=0$). 

Below, we compute $R$ at depinning both for short range (respectively
long range) elasticity to lowest non-trivial order in $\epsilon = 4-d$
(respectively $\epsilon = 2-d$), i.e within a 1-loop
calculation. However, since it turns out that the momentum integrals
involved in the calculation depend very strongly on the dimension, we
found it useful, and sometimes more accurate, to carry a 1-loop
approximation directly in fixed dimension $d$. Also, since there is
one exact result for a massive propagator, we  also give the
result in that case.

We denote:
\begin{equation}\label{lf10}
 g(q) = \frac{1}{q^2}
\end{equation}
with the obvious change for long range elasticity $g(q) = \frac{1}{|q|}$,
and (see below) the massive propagator. 

The final result in the continuum is given by the sum of the
two diagrams in Fig.~\ref{fig3}:
\begin{eqnarray} \label{lf22a}
D&=& \int_{xy} \left< u_{xt}^2 u_{yt}^2 \right>_c\nonumber \\
& =& -
2 \Delta'(0^+)^4 L^{d}
\int \frac{\rmd^d q}{(2 \pi)^d} \frac{\rmd^d k}{(2 \pi)^d} \frac{\rmd^d p}{(2
\pi)^d}\times \nonumber \\
&& \Big[ 2 g(q)^2 g(k)^2 g(p)^2 g(p+q) g(p+k)\nonumber  \\
&& +
  g(q)^2 g(k)^2 g(p) g(p+q) g(p+k) g(p+k+q) \Big] \nonumber \\
\end{eqnarray}
\begin{equation}\label{lf11}
\int_{xy} G_{xy}^2 = L^d \int \frac{\rmd^d q}{(2 \pi)^d}
G(q)^2 = L^d \Delta(0)^2  \int \frac{\rmd^d q}{(2 \pi)^d} g(q)^4
\end{equation}
The combinatorics can be done as follows. There is a factor $1/ ( 4!
2^4)$. There are $4!$ ways to associate each one of the four external
$u$ to an unsplitted vertex. Say $1,2$ are now linked to $u_x^2$ and
$3,4$ to $u_y^2$. At each vertex a $\hat u$ field comes out. There are
$2^4$ ways to assign them to each splitted vertex. Then there is a
unique set of four splitted points, one at each vertex, entering the
acausal graph (which - in effect - is the only one arising, with the
minus sign).  But there are still 3 ways to join these four points in
a loop: Two give the first integral, one the second, and finally for
each case the orientation can be chosen in two ways.

Let us go to the discrete model with periodic BC \cite{caveat}. We recall that
\begin{equation}\label{lf12}
 L^d \int \frac{\rmd^d q}{(2 \pi)^d} f(q) \equiv \sum_{n \in \mathbb{Z}^d}
f\!\left(\frac{2 \pi n}{L}\right)\ ,
\end{equation}
where here and in the following the term with $n= 0$ is always
excluded. 
In the limit of large $L/a$ one finds:
\begin{eqnarray} \label{lf25}
D&=& a^{2 d} \sum_{xy} \left< u_{xt}^2 u_{yt}^2 \right>_c  \\
&=& -
2 \Delta'(0^+)^4 L^{- 2 d} (\frac{L}{2 \pi})^{16} \times \nonumber \\
&& \sum_{n,m,l \in \mathbb{Z}^d}
\bigg[ 2 \frac{1}{(n^2)^2 (m^2)^2 (l^2)^2 (l+n)^2 (l+m)^2 }\nonumber
\\
&& \hphantom{\sum }+  \frac{1}{(n^2)^2 (m^2)^2 (l^2)
(l+n)^2 (l+m)^2 (l+m+n)^2} \bigg]\nonumber  \\
&& a^{2 d} \sum_{xy} G_{xy}^2 = \Delta(0)^2 \left(\frac{L}{2 \pi}\right)^{8}
\sum_{n \in \mathbb{Z}^d} \frac{1}{n^8}
\label{lf26}
\end{eqnarray}

One can see that the ratio $R$ will be universal
since the 1-loop FRG fixed point equation taken 
at $u=0$ yields:
\begin{eqnarray}\label{lf27}
 (\epsilon - 2 \zeta) \Delta(0) &=& (\epsilon I)  \Delta'(0^+)^2 \\
 (\epsilon I) = L \partial_L I &=& \frac{2}{(4 \pi)^{d/2} \Gamma (
3-\frac{d}{2})}  \label{lf28}
= \frac{1}{8 \pi^2}\qquad \ \
\end{eqnarray}
The last identity is valid for $d=4$. In fact, since the 1-loop FRG
equation is universal, it holds as well for $d=4$ as for $d<4$. For
$d<4$, we use
\begin{eqnarray}\label{lf29}
I &=& \int \frac{\rmd ^d p}{(2 \pi)^d} \frac{1}{p^4}
\equiv L^{-d} \sum_{n \in \mathbb{Z}^d} \frac{1}{(2 \pi n/L)^4}\nonumber \\
&=& \frac{L^\epsilon}{ (2 \pi)^{4}} \sum_{n \in \mathbb{Z}^d} \frac{1}{n^4} \\
L \partial_L I &=& \epsilon I = \epsilon \frac{L^\epsilon}{ (2 \pi)^{4}}
\sum_{n \in \mathbb{Z}^d} \frac{1}{n^4} \label{lf30}
\end{eqnarray}
This is all we need to compute the universal ratio.

\subsection{Calculation to lowest order in $\epsilon=4-d$}
The ratio $R=D/(2 \sum_{xy} G_{xy}^2)$ is:
\begin{eqnarray}\label{lf31}
 R &=& - \epsilon^2 (1 - 2 \zeta_1)^2
(8 \pi^2)^2 \frac{1}{(2 \pi)^8}
\frac1{\sum_{n \in \mathbb{Z}^d} \frac{1}{n^8}} \times \nonumber \\
&& \sum_{n,m,l \in \mathbb{Z}^d}
\bigg[ 2 \frac{1}{(n^2)^2 (m^2)^2 (l^2)^2 (l+n)^2 (l+m)^2 }\nonumber
\\
&& +  \frac{1}{(n^2)^2 (m^2)^2 (l^2) (l+n)^2 (l+m)^2 (l+m+n)^2}
\bigg]\nonumber  \\
\end{eqnarray}
One finds, using
\begin{equation}\label{lf13}
\sum_{n \in \mathbb{Z}} \rme^{- t n^2} = \Theta(3,0,\rme^{-t})
\end{equation}
that in $d=4$
\begin{equation}\label{lf14}
\sum_{n \in \mathbb{Z}^d} \frac{1}{n^8} = \frac{1}{6} \int_0^\infty
t^3 (\Theta(3,0,\rme^{-t})^d - 1) = 10.2454
\end{equation}
Noting $f(p) = \sum_{q \in \mathbb{Z}^d} \frac{1}{(q^2)^2 (p+q)^2}$ one has:
\begin{eqnarray}\label{lf33}
&& \!\!\!\!\!\!\!\!\!\!\!\!\!\!\!\!\!\!  \sum_{q,k,p \in \mathbb{Z}^d}
\frac{1}{(q^2)^2 (k^2)^2
(p^2)^2 (p+k)^2 (p+q)^2}
\nonumber \\
&=& \sum_{p \in \mathbb{Z}^d} \frac{1}{(p^2)^2} f(p)^2 \approx  1850\ . \\
&& \!\!\!\!\!\!\!\!\!\!\!\!\!\!\!\!\!\! \sum_{n,m,l \in \mathbb{Z}^d}
\frac{1}{(n^2)^2 (m^2)^2
(l^2) (l+n)^2 (l+m)^2 (l+m+n)^2} \nonumber \\
&\approx& 980. \label{lf34}
\end{eqnarray}
The final result is:
\begin{equation}\label{lf15}
 R = - 1.17 \frac{1}{9} \epsilon^2  \approx - 0.13  \epsilon^2
\end{equation}
This results shows that $R$ is quite small near $d=4$,
but increases quite fast as the dimension is lowered. However
the sums over the momenta  depend very strongly on $d$ (see below)
and one should expect significant higher order corrections
in $\epsilon$. Thus the result (\ref{lf15}) is likely to drastically 
overestimate the (absolute value of the)
result in lower dimensions, which is why we now turn to
an estimate in fixed dimension. 

\subsection{1-loop estimate in general dimension}
One can perform an estimate in general dimension, based on
an arbitrary truncation on (i) 1-loop graphs (ii) neglect of
the finite size scaling function.

In general dimension one has the dimensionless ratio:
\begin{eqnarray}
 R &=& - (\epsilon - 2 \zeta)^2
\frac1{(\epsilon \sum_{n \in \mathbb{Z}^d} \frac{1}{n^4} )^2 \sum_{n \in \mathbb{Z}^d}
\frac{1}{n^8}} \nonumber \\
&& \sum_{n,m,l \in \mathbb{Z}^d}
\bigg [  \frac{2}{(n^2)^2 (m^2)^2 (l^2)^2 (l+n)^2 (l+m)^2 }\nonumber
\\
&& +  \frac{1}{(n^2)^2 (m^2)^2 (l^2) (l+n)^2 (l+m)^2 (l+m+n)^2} \bigg
]\nonumber \\  \label{lf35}
\end{eqnarray}
Let us  give a table of values:
\begin{equation}
\begin{array}{rll}
\displaystyle \sum_{n \in \mathbb{Z}^d} \frac{1}{n^4} &= \displaystyle
\frac{\pi^4}{45} =
2.16465 \qquad&
\displaystyle ( d=1) \\
&\displaystyle = 6.0268120 \qquad  & ( d=2) \\
& \displaystyle= 16.5 \qquad & (d=3)
\end{array} \label{lf36}
\end{equation}
\begin{equation} \label{lf37}
\begin{array}{rll}
\displaystyle \sum_{n \in \mathbb{Z}^d} \frac{1}{n^8} &=\displaystyle
\frac{\pi^8}{4725}  = 2.00815 &\displaystyle \qquad ( d=1) \\
& \displaystyle = 4.28143066080578& \displaystyle\qquad  ( d=2) \\
& \displaystyle= 6.9458079272  &\displaystyle\qquad ( d=3 )
\end{array}
\end{equation}
\begin{equation}
\begin{array}{rll}
 \displaystyle\sum_{n,m,l \in \mathbb{Z}^d}&\displaystyle  \frac{1}{(n^2)^2 (m^2)^2 (l^2)^2 (l+n)^2 (l+m)^2 } \nonumber \\
& \displaystyle= 0.37342751117 &( d=1) \\
& \displaystyle = 26.567 & (d=2) \\
& \displaystyle = 240 &(d=3 )
\end{array} \label{lf38}
\end{equation}
\begin{equation}
\begin{array}{rll}
\displaystyle \sum_{n,m,l \in \mathbb{Z}^d}&\displaystyle \frac{1}{(n^2)^2
(m^2)^2 (l^2) (l+n)^2 (l+m)^2
(l+m+n)^2} \hspace{-2cm}&\hspace{2cm} \\
&= 0.069672062794960 & ( d=1) \\
& = 14.138 &  (d=2) \\
& = 143 & ( d=3)
\end{array} \label{lf39}
\end{equation}

In $d=1$ one finds:
\begin{eqnarray} \label{lf40}
 R &=& - 0.0867759691287\left (1 - 2 \frac{\zeta}{\epsilon}\right)^2 \\
&  \approx& - 0.00964177 \qquad (\zeta=\epsilon/3)  \label{lf41}\ \\
&  \approx& - 0.00347       \hphantom{177}      \qquad (\zeta=1.2)
\label{lf42} 
\end{eqnarray}
In $d=2$ one finds:
\begin{eqnarray}
 R &=& - 0.4326 \left(1 - 2 \frac{\zeta}{\epsilon}\right)^2 \nonumber \\
  &\approx& - 0.0481 \qquad (\zeta=\epsilon/3)  \label{lf43}
\end{eqnarray}
In $d=3$ one finds:
\begin{eqnarray}
 R &=& - 0.3297 \left(1 - 2 \frac{\zeta}{\epsilon}\right)^2 \nonumber \\
&\approx& - 0.0366 \qquad (\zeta=\epsilon/3)  \label{lf44}
\end{eqnarray}
where we have inserted various choices for $\zeta$ including the
1-loop result $\zeta=\epsilon/3$. One finds that already
in $d=3$ the 1-loop approximation is significantly
lower than the extrapolation from (\ref{lf15}) as discussed above.

\subsection{Long range elasticity in general dimension}

For LR elasticity the upper critical dimension is $d_{\mathrm{uc}}=2$.
The general expression for $R$ is:

\begin{eqnarray}
 R &=& - (\epsilon - 2 \zeta)^2
\frac1{(\epsilon \sum_{n \in \mathbb{Z}^d} \frac{1}{n^2} )^2 \sum_{n \in \mathbb{Z}^d}
\frac{1}{n^4}} \nonumber \times \\
&& \sum_{n,m,l \in \mathbb{Z}^d}
\bigg[  \frac{2}{(n^2) (m^2) (l^2) |l+n| |l+m| }\nonumber
\\
&& \qquad \qquad +  \frac{1}{(n^2) (m^2) |l| |l+n| |l+m| |l+m+n|} \bigg ]\nonumber \\ \label{lf45}
\end{eqnarray}

It is interesting to compute $R$ in $d=1$. Using that:
\begin{eqnarray}
&& \sum_{n \in Z} \frac{1}{n^2} =  \frac{\pi^2}{3} = 3.28987 \label{lf46} \\
&& \sum_{n \in Z} \frac{1}{n^4} = \frac{\pi^4}{45} = 2.16465  \label{lf47}\\
&& \sum_{n,m,l \in Z} \frac{1}{(n^2) (m^2) (l^2) |l+n| |l+m| }= 3.847  \label{lf48}\\
&& \sum_{n,m,l \in Z} \frac{1}{(n^2) (m^2) |l| |l+n| |l+m| |l+m+n|} =
1.934 \nonumber \\  \label{lf49}
\end{eqnarray}
one finds:
\begin{eqnarray}
R &=& - 0.4109\left (1 - 2 \frac{\zeta}{\epsilon}\right)^2  \nonumber
\label{lf50} \\ 
&\approx& - 0.04566 \qquad (\zeta=\epsilon/3) \label{lf51}
\end{eqnarray}

\subsection{Long range epsilon expansion}

Similarly, one can perform an expansion in $\epsilon=2-d$. 
In $d=2$ one has:
\begin{eqnarray}
&& \epsilon I = 1/(2 \pi)  \label{lf52} \\
&& \sum_{n \in Z^2} \frac{1}{n^4} = \frac{\pi^4}{45} = 6.0268120
\label{lf53} \\ 
&& \sum_{n,m,l \in Z^2} \frac{1}{(n^2) (m^2) (l^2) |l+n| |l+m| } = 550
\pm 20  \label{lf54}\\
&& \sum_{n,m,l \in Z^2} \frac{1}{(n^2) (m^2) |l| |l+n| |l+m|
|l+m+n|} = 370 \pm 10 \nonumber \\   \label{lf55}
\end{eqnarray}
This yields
\begin{eqnarray}
R &=& - \epsilon^2 \left(1 - 2 \zeta_1\right)^2
(2 \pi)^2 \frac{1}{(2 \pi)^4}
\frac1{\sum_{n \in \mathbb{Z}^d} \frac{1}{n^4}} \times \nonumber\\
&& \sum_{n,m,l \in \mathbb{Z}^d}
\bigg [ 2 \frac{1}{(n^2) (m^2) (l^2) |l+n| |l+m| }\nonumber
\\
&& +  \frac{1}{(n^2) (m^2) |l| |l+n| |l+m| |l+m+n|} \bigg ] \qquad
\label{lf56} 
\end{eqnarray}
The result is 
\begin{eqnarray}
 R &=& - 6.17 \frac{1}{9} \epsilon^2 \approx - 0.686 \epsilon^2 \label{lf57}
\end{eqnarray}
again, a probable overestimation of the result if naively extrapolated
to $d=1$. 

\subsection{Harmonic well, SR elasticity}

It is interesting to compare with the calculation in
a massive scheme, i.e.\ an interface in a harmonic well.
Setting $g(q):=1/(1+q^2)$, we have for $d=1$
\begin{eqnarray}
&&\int_q g(q)^2 = \frac \pi 2 = 1.5708  \label{lf58}\\
&&\int_q g(q)^4 = \frac {5 \pi} {16} = 1.62596  \label{lf59}\\
&&\int_q \int_k\int_p  g(q)^2 g(k)^2 g(p)^2 g(p+q) g(p+k)\nonumber \\
 \label{lf60}
&&\qquad =
\frac{1631\pi^3}{31104} = 1.62588  \label{lf61}\\
&&\int_q \int_k\int_p  g(q)^2 g(k)^2 g(p) g(p+q) g(p+k) g(p+k+q) \nonumber \\
&&\qquad = \frac{245 \pi^3}{5184} = 1.46538 \label{lf62}
\end{eqnarray}
This gives the ratio
\begin{equation}\label{lf63}
R= - \frac{2366}{1215}  (1-2 \zeta/\epsilon )^2 = - \frac{2366}{10935}
= - 0.216369455  \label{lf64}
\end{equation}

It is interesting to compare the present result to one case (to our
knowledge the only one apart from mean field models) where the {\it
full distribution of $u$} is known in a non-trivial disordered problem
\cite{LeDoussalMonthus}.  This is the {\it static} random field model
in $d=0$ in a harmonic well (i.e.\ the massive case), the so-called
toy model. The exact result there is:
\begin{equation}\label{lf63a}
R_{\mathrm{toy}} = -0.080865\dots 
\end{equation}
The present 1-loop approximation for the problem of depinning,
continued to $d=0$ would give the larger result 
 $R=-1/3$. It is unclear at present whether the
difference between the two results indicates that the
1-loop approximation is unsatisfactory so far from
$d=4$, or if statics and depinning have radically
different values of $R$.

\section{Behavior at the critical dimension} \label{sec5} In this
Section we reexamine isotropic depinning, and statics, exactly in
$d=4$. We solve the RG equations in $d=4$ and obtain the behavior of
the correlation function. Contrarily to periodic systems at the upper
critical dimension \cite{ChitraGiamarchiLe1999}, non-periodic objects
such as interfaces submitted to either random bond or random field
type disorder exhibit a roughness, which is a power of a logarithm.

The FRG flow equation ($\beta$-function) for the (renormalized) force
correlator has a good limit for $d=4$. If one defines
\begin{equation}\label{lf66}
\Delta_l(u) = 8 \pi^2 l^{2 \zeta_1-1} \tilde \Delta_l(u l^{- \zeta_1})
\end{equation}
with $l=\ln(\Lambda/m)$ ($\Lambda$ some UV cutoff) then 
the function $\tilde \Delta_l(u)$ satisfies
\begin{eqnarray}\label{rg2}
 \partial_l {\tilde \Delta}(u) &=& (1 - 2 \zeta_1) \tilde \Delta(u) 
+ \zeta_1 u \tilde \Delta'(u) \nonumber \\
&& - \frac{1}{2}
\left[(\tilde \Delta(u) - \tilde \Delta(0))^2 \right]''
+ l^{-1} \beta_2(\tilde{\Delta})
\label{rg2a}\ .
\end{eqnarray}
where, for depinning,
\begin{eqnarray}
\beta_2(\Delta)  &=& \left[ (\Delta(u) - \Delta(0)) \Delta'(u)^2 \right]''
+ \Delta'(0^+)^2 \Delta''(u)
\ . \nonumber
\end{eqnarray}
and $\zeta_1= \zeta/\epsilon=1/3$ is the 1-loop value, see e.g.\
\cite{LeDoussalWieseChauve2002}.  It is then easy to see that the
function $\tilde \Delta_l(u)$ converges towards the 1-loop fixed point
with the following asymptotic corrections:
\begin{equation}\label{lf16}
 \tilde \Delta_l(u) = \tilde \Delta^*(u) +
\sum_n l^{- \omega_n} b_n(u) + \frac{1}{l} \beta_1'[\tilde{\Delta}^*]^{-1}
\beta_2(\tilde{\Delta}^*)
\end{equation}
where $(\beta_1')^{-1}$ is the inverse of the linearized 1-loop
$\beta$-function and the $\omega_n$ are the 1-loop eigenvalues.

Using (\ref{lf66}) with $\zeta_{1}=1/3$ yields the result for the
correlation function at $q=0$:
\begin{eqnarray}
 \langle u_q u_{-q} \rangle|_{q \ll m}  &=& m^{-4} \Delta(0) \\ 
&=& c  m^{-4} \ln(\Lambda/m)^{- \frac{1}{3}}
(1 + O(1/\ln(\Lambda/m)) )\nonumber 
\end{eqnarray}
as $m\to 0$, with $c=8 \pi^2 \tilde \Delta^*(0)$, both for statics and
depinning; the difference lies in the subdominant piece. Within the
present approach using the renormalization scheme at $q=0$, the
2-point correlation function at non-zero $q$ can be computed from the
renormalized (uniform) effective action by resumming an infinite set
of diagrams. Using the standard finite size scaling ansatz allows to
obtain the other limit of the scaling function, where $\Lambda \gg q
\gg m$.  To lowest order (one loop) in the renormalized disorder one
has \cite{LeDoussalWieseChauve2002a}
\begin{eqnarray}
 (q^2 + m^2)^2 \langle u_q u_{-q} \rangle & = &
( \Delta(0) -
\Delta'(0^+)^2 (I(q) - I(0))  + \dots  ) \nonumber  \\
I(q)&=&\int_p \frac1{(p^2 + m^2)((p+q)^2 + m^2)}
\ .  \label{lf65}
\end{eqnarray}
This gives:
\begin{equation}
 \langle u_q u_{-q} \rangle = c (q^2 + m^2)^{-2} \,
\ln\!\left(\frac{\Lambda}{m}\right)^{\!\!- \frac{1}{3}} \left( 1 -
\frac{1}{3} \frac{\ln(\frac{m}{q})}{\ln(\frac{\Lambda}{m})} + \dots
\right)
\end{equation}
Assuming scaling, i.e.\ that the function
$( 1 - \frac{1}{3} x+\dots ) \to (1 + x)^{-\frac{1}{3}}$ one finds
that:
\begin{equation}
 \langle u_q u_{-q} \rangle \sim q^{-4} \left[\ln(\textstyle
\frac{m}{q})\right]^{-1/3}
\end{equation}
and thus:
\begin{equation}
  \overline{(u_x - u_0)^2}  \sim (\ln x)^{2/3}
\end{equation}

\section{Stability of the 1-loop fixed point} \label{sec6} Here we
analyze the stability of a functional fixed point. Two cases have to
be distinguished:
\begin{itemize}
\item [(a)] There is the freedom to rescale
the field $u$ while at the same time rescaling the disorder
correlator. This includes the random bond and random field interface
models. 
\item [(b)] There is no such freedom, since the period is fixed by the
microscopic disorder. This is the case for a charge density wave
(random periodic problem), but also for the random field bulk problem,
in its treatment via a non-linear sigma model.
\end{itemize}
We first analyze the simpler case (b).

\subsection{Stability of a functional fixed point; periodic case}\label{(b)}
Be the flow-equation given by
\begin{equation}
\partial_{l} R (u) = \beta[R] (u) = \epsilon R (u) + f[R,R] (u)\ ,
\end{equation}
where $f$ is some bilinear form of $R$, which contains at least one
derivative for each $R$.  A similar equation of
course exists for $\Delta (u)= -R'' (u)$ and the corresponding 
$\beta[\Delta](u)$. 

Suppose $R^{*} (u)$ is the non-trivial fixed point of order
$\epsilon$, i.e.\ $\beta [R^{*}]=0$.  Two eigenfunctions and
eigenvalues above $R^{*} (u)$ can be identified. \begin{enumerate}
\item [(i)] The constant mode $\delta R (u)= 1$ with eigen-value
$\epsilon$. (As long as it is permissible physically.)
\item [(ii)] The first subleading eigenfunction $\delta R (u)= R^*(u)$
with eigenvalue $-\epsilon$.
\end{enumerate}
\leftline{\underline{Proof:}}
For case (i), we have for $\kappa \ll 1$
\begin{equation}
\partial_{l} \left ( R^{*} (u)+\kappa \right)  = \beta [R^{*}+\kappa]
(u) = \epsilon \kappa \ ,
\end{equation}
since $f$ does not couple to the constant by assumption. 
This proves (i).

For case (ii), we have ($\kappa \ll 1$):
\begin{eqnarray}
&&\!\!\!\partial_{l} \left ( R^{*} (u)+\kappa R^{*} (u) \right)  = \beta
[R^{*} (1+\kappa)] (u) \nonumber \\
&&\qquad = \epsilon R^{*} (u) (1+\kappa) + (1+\kappa)^{2} f [R^{*},R^{*}] (u)
\end{eqnarray}
Subtracting $\beta [R^{*}] (u)=0$ on the r.h.s.\ and expanding for small $\kappa$ yields
\begin{eqnarray}
\partial_{l} \left ( R^{*} (u)+\kappa R^{*} (u) \right) &=& \epsilon
\kappa R^{*} (u) + 2 \kappa f [R^{*},R^{*}] (u) \nonumber \\
&=& - \epsilon \kappa R^{*} (u)\ ,
\end{eqnarray}
where in the last equation we have again used the fixed-point
condition $\beta [R^{*}] (u)=0$, i.e.\ $\epsilon R^{*} (u) +
f[R^{*},R^{*}] (u)=0$, to eliminate $f[R^{*},R^{*}] (u)$. This proves
(ii).

Using the same line of arguments, it is easy to see that when starting
from $R (u)\sim R^{*} (u)$ with some arbitrary amplitude, the flow is
always remaining on the critical manifold spanned by $R^{*} (u)$.

Of course, there are in general more eigen-functions and
eigen-values. See \cite{LeDoussalWieseChauve2002} for an explicit
example.

\subsection{Perturbations of the fixed
point in presence of the freedom to rescale}\label{gentheoperfixpt}
We state the following
\smallskip

\leftline{\underline{Theorem:}}\noindent 
The differential equation of the form
\begin{eqnarray}\label{t0}
-m \partial_{m} \Delta (u) &=& \beta [\Delta] \nonumber \\
\beta [\Delta] 
&=& (\epsilon -2\zeta)\Delta (u) + \zeta u
\Delta' (u) + f [\Delta,\Delta]\qquad  \label{t1}
\end{eqnarray}
where the symmetric functional $f [\Delta,\Delta ]$ transforms under
$\Delta (u) \to \kappa^{-2}\Delta (\kappa u)$ in the same way as
$\Delta$, has the two eigenfunctions and eigenvalues of perturbations around
the fixed point $\beta [\Delta^{*}]=0$
\begin{eqnarray}\label{t2.0}
z_{0} (u) &=& u \Delta' (u) -2 \Delta (u) \\
\lambda_{0} &=& 0 \ . \label{lf70.0a}\\
\label{t2}
z_{1} (u) &=& \zeta u \Delta' (u) + (\epsilon -2\zeta) \Delta (u) \\
\lambda_{1} &=& -\epsilon \ . \label{lf70.a}
\end{eqnarray}
(We note $\Delta$ instead of $\Delta^{*}$ for the fixed point for
simplicity of notations.)  Note that the assumptions are satisfied by
the 1-loop flow-equation (RF-case).  \smallskip

\leftline{\underline{Proof:}}\noindent 
Be $\beta [\Delta ] (u) =0$.  
Due to the assumptions, for all $\kappa$
\begin{equation}\label{t3}
\beta [ \kappa ^{-2} \Delta]  (\kappa u)= 0\ .
\end{equation}
Deriving w.r.t.\ $\kappa$ gives with (\ref{t1}) at $\kappa =1$
\begin{eqnarray}\label{t4}
&&  (\epsilon
-2 \zeta) \left(u \Delta' (u)-2 \Delta (u) \right) + \zeta u \left(u
\Delta' (u)-2 \Delta (u) \right)' \nonumber \\
& &\qquad + 2 f \left( \Delta (u), 
u \Delta' (u)-2 \Delta (u) \right) =0 \ \ \ 
\end{eqnarray}
This is nothing but the eigenvalue equation for the perturbation $z_{0
} (u)$ about the fixed point $\beta [\Delta]=0$ and proves the
existence of the solution $z_{0} (u)$ with eigenvalue
$\lambda_{0}=0$. Note that this {\em redundant} operator with
eigenvalue 0 persists to all orders in perturbation theory.

Let us turn to the next solution.
Multiplying (\ref{t4}) with $\zeta$ and adding $2\epsilon$ times
$\beta[\Delta] =0$ gives
\begin{eqnarray}\label{t5}
&&(\epsilon -2 \zeta) \left(\zeta u \Delta' (u)+( 2 \epsilon -
2\zeta) \Delta (u) \right) \nonumber\\ 
& &\qquad +\, \zeta u \left(\zeta u \Delta' (u)+ (2 \epsilon -2\zeta) \Delta(u) \right)' \nonumber \\
&&\qquad +\, 2 f \left( \Delta (u), \zeta u
\Delta' (u)+ (\epsilon -2\zeta) \Delta (u) \right)=0\ ,\quad \qquad 
\end{eqnarray}
where we used the bilinearity of $\Delta (u)$. 
Rearranging yields
\begin{eqnarray}\label{t6}
&&(\epsilon -2 \zeta) \left(\zeta u \Delta' (u)+( \epsilon -
2\zeta) \Delta (u) \right) \nonumber\\ 
& &\qquad +\, \zeta u \left(\zeta u \Delta' (u)+ (\epsilon -2\zeta) \Delta(u) \right)' \nonumber \\
&&\qquad +\, 2 f \left( \Delta (u), \zeta u
\Delta' (u)+ (\epsilon -2\zeta) \Delta (u) \right) \nonumber \\
&&\qquad = -\epsilon (\zeta u \Delta' (u)+(\epsilon - 2\zeta)\Delta (u))\ .
\end{eqnarray}
This equation is nothing but the eigenvalue-equation
for the perturbation $z_{1} (u)=\zeta u \Delta' (u)+ (\epsilon -2\zeta)\Delta
(u)$ about the fixed point $\beta [\Delta]=0$ with eigen-value
$\lambda_{1}=-\epsilon$. q.e.d.
\smallskip

\leftline{\underline{Remark:}}\noindent Consider the case of
short-range disorder, i.e.\ that $\Delta (u)$ falls off rapidly, and
is monotonic. Usually, the leading fixed point solution has no knot
(no $u$ such that $\Delta (u)=0$). Then $z_{0} (u)$ has no knot and
$z_{1} (u)$ has one knot. Eigenvalues should be ordered (like in
quantum mechanics) due to their number of knots. Thus we should have
found 
the two leading  solutions for short-range disorder. This is confirmed
by the numerical analysis given in  subsection \ref{stability:numeric}. 

\medskip 

\leftline{\underline{Corollary: (Random bond case)}}\smallskip

\noindent 
The differential equation of the form
\begin{eqnarray}\label{t0p}
-m \partial_{m} R (u) &=& \beta [R] (u)  \\
\beta [R] (u) 
&=& (\epsilon -4\zeta)R (u) + \zeta u
R' (u) + f [R,R] (u)\nonumber \qquad 
\end{eqnarray}
where the symmetric functional $f [R,R ]$ transforms under
$R (u) \to \kappa^{-4}R(\kappa u)$ in the same way as
$R$, has the two eigenfunctions and eigenvalues of perturbations around
the fixed point $\beta [R^{*}]=0$
\begin{eqnarray}\label{t2.0a}
z_{0} (u) &=& u R' (u) -4R (u) \\
\lambda_{0} &=& 0 \ . \label{lf70.0}\\
\label{t2a}
z_{1} (u) &=& \zeta u R' (u) + (\epsilon -4\zeta) R (u) \\
\lambda_{1} &=& -\epsilon \ . \label{lf70}
\end{eqnarray}
Note that the assumptions are satisfied by the 1-loop flow-equation
(random bond case). 

\medskip \leftline{\underline{Proof:}}
\smallskip
This can either be proven along the same lines as for (\ref{t1}) or by
deriving $z_{0} (u)$ and $z_{1} (u)$ twice w.r.t.\ $u$ and then using
the theorem for the random field case (\ref{t1}).

\subsection{Numerical analysis of the RF-fixed point}\label{stability:numeric}
We start from the 1-loop flow equation
\begin{eqnarray}
-m \partial_m {\Delta}(u) &=&  (\epsilon - 2 \zeta) \Delta(u)
+ \zeta u \Delta'(u) \nn\\
&& - \frac{1}{2} \left[(\Delta(u) - \Delta(0))^2\right]'' 
\label{rgdisorder1l}\ .
\end{eqnarray}
It has  the following solution \cite{DSFisher1986,LeDoussalWieseChauve2002} 
\begin{eqnarray}\label{sol}
\Delta (u) &=& \frac{\epsilon}{3} y_{1}(u) \nn \\
y_{1} (u)-\ln y_{1} (u)&=& 1+ \frac{1}{2} u^{2} \ .
\end{eqnarray}
Perturbations around this solution satisfy the differential equation
\begin{eqnarray}
&&\!\!\!-m \partial_m \left[{\Delta}(u) + z (u) \right]= \lambda
\epsilon z(u) \label{lf67} \\ 
&&\!\!\! \left(1-3 \lambda \right)z (u) +  u z' (u)- 
\left[(y_{1}(u) - y_{1}(0)) (z (u)-z (0))\right]''\nonumber \\
&& \qquad =0 
\label{rgdisorder1la}
\end{eqnarray}
In order to have a criterion for the numerical integration, one has to
determine the behavior at infinity. Using that for $u \to \infty $ 
\begin{equation}\label{asympy1}
y_1 (u) \approx  \rme^{-1-u^{2}/2}\ ,
\end{equation}
and assuming exponentially fast decay for $z(u)$,
one finds that the asymptotic  behavior is 
\begin{equation} \label{asymptot}
z (u) \approx  z (0) \, \frac{(u^{2}-1) \rme^{-1-u^{2}/2}}{2+3\lambda } \ .
\end{equation}
Quite surprisingly, the asymptotic behavior is fixed, including
its amplitude\footnote{We thank O.~Narayan for this observation.}.
In any case, slower power law decay is ruled out on
physical ground since we are considering short
range disorder. 

The solutions for $\lambda =0$ and $\lambda =-\epsilon$ are given in
Eqs.~(\ref{t2.0}) ff.
\begin{figure}[t]
{\unitlength0.01\columnwidth
\begin{picture} (100,60)
\put(0,0){\Fig{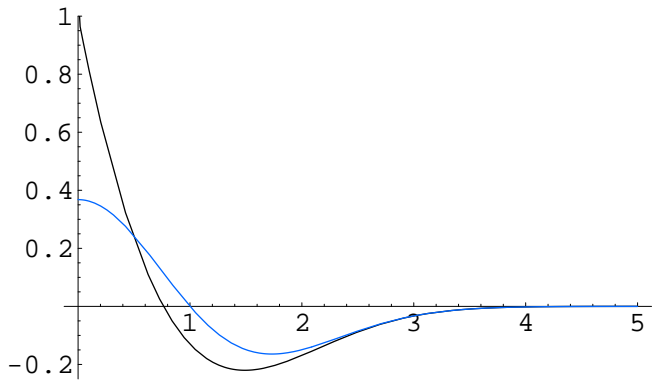}}
\end{picture}
}
\caption{The  solution $z_{1} (u)$ (in black, with $z (0)=1$) and
asymptotic behavior as given in (\ref{asymptot}) (blue/bright).}
\label{f:pert}
\end{figure}
 The solution $z_{1} (u)$ is dominant, and gives the correction to
scaling exponent $\omega = - \epsilon$. Note that this exponent is
minus the engineering dimension of the coupling, as is the case in
standard field-theory\cite{Zinn}, and also for the random periodic
class \cite{LeDoussalWieseChauve2002}.  Ref. \cite{RamanathanFisher1998}
cites the value $\omega =- \epsilon/3$. We find the corresponding
numerical solution to decay as $u^{-2}$, incompatible with
(\ref{asymptot}) and physically unacceptable [34]. 

The question arises, of whether there are more solutions with fast
decay. Intuitively,
one would expect this: Making $\lambda$ more negative, the solution
overshoots and one might think of fine-tuning $\lambda$ such that it
approaches the axis for large $u$ from above. However this is
incompatible with the asymptotic behavior in Eq.~(\ref{asymptot}),
which predicts, that all solutions for $\lambda <-2/3$ 
converge from below. In fact, we  have not been able to find any
further solution, and we conjecture that there is none. It would be
interesting to prove this rigorously. This behavior is in contrast to the
random periodic case, solved in \cite{LeDoussalWieseChauve2002},
which has  infinite many subleading contributions.

\section{Conclusion}

In this paper we have explored further properties of the field theory
of depinning. We have defined and computed universal observables, such
as the distribution of the interface width and the ratio $R$ of the
connected 4-point cumulant to the square of the 2-point one
(kurtosis).  This ratio measures the deviations from a Gaussian
approximation which we have also used to obtain the universal
distribution.  Higher order connected cumulants can be computed in a
similar way to one loop using polygon diagrams, and one should be able
to reconstruct the full distribution from them. Other properties of
the theory such as the behavior at the upper critical dimension and
the finite size scaling behavior have been clarified.  All
calculations in the present paper are of interest for comparison with
numerical simulations, existing ones
\cite{RossoKrauthLeDoussalVannimenusWiese2002} or in the near future.

In the process of computing the 4-point function we discovered massive
cancellations between diagrams. We traced this back to the physically
expected property that correlations exactly at depinning should be
time independent. A diagrammatic proof of this property is still
incomplete, but we have provided some convincing elements in that
directions. As a result the correlation functions can be computed in a
much simpler way. Thus there seems to be an underlying theory, with
``quasi-static'' diagrams (i.e.\ not containing time explicitly), with
some additional rules. We have understood these rules to lowest
(1-loop) order and it would be of high interest to understand --and
prove-- them to all orders. It is even possible that there exists a
simpler formulation of the theory at depinning in terms of,
e.g.\ effective fermions. The fermionic character is suggested by the
cancellation of all diagrams except for the ``acausal loops'' with a
minus sign. We thus encourage further examination of this fascinating
question and full elucidation of the field theory which describes
depinning.

\begin{appendix}
\def\theequation{\thesection.\arabic{equation}}

\section{Different boundary conditions}\label{differentbound}

\subsubsection{Periodic boundary conditions}\label{periodicbound}
A periodic function $u$ with period $L$ satisfies
\begin{equation}\label{lf71}
u(x+L \hat{e}_{i} ) = u(x) \ ,
\end{equation}
where the $\hat{e}_{i}$ are orthonormal.
It can be written as
\begin{equation}\label{lf72}
u (x) = \sum \tilde u_{ k}\, \rme^{ikx} \ ,
\end{equation}
where summation runs over all $k$, such that $k_{i}=n_{i} 2\pi /L$,
$n_{i}\in \mathbb{Z}$.

\subsubsection{Open boundary conditions}\label{openbound}
To simplify the notation, we give all formulas for one dimension;
generalizations are straightforward.

Suppose the function $f (x)$ is defined on $[0,L]$. Then a function $g
(x)$ can be defined by the following prescription
\begin{equation}\label{lf80}
g (x) = \left\{\begin{array}{lcl}
f (x)&\mbox{  for  }& 0\le x\le L \\
f (2L-x)& \mbox{  for  }& L<x\le 2L
\end{array} \right.
\end{equation}
$g (x)$ can be prolonged to a periodic function with period $2 L$,
i.e.\ $g (x+2L)=g (x)$, and satisfies by construction in addition
\begin{equation}\label{addcons}
g(2L-x)=g (x)\ .
\end{equation}
In the basis needed to construct
functions with period $2L$, we have to restrain ourselves to
\begin{equation}\label{lf81}
g(x) =  \sum \tilde g_{n} \cos \left(\frac{2 \pi n
x}{2L} \right)=  \sum \tilde g_{n} \cos \left(\frac{ \pi n x}{L} 
\right)\ ,
\end{equation}
since the $\sin$ do not satisfy (\ref{addcons}). The such constructed
set of functions $f (x)$, $x\in [0,L]$ has Neumann-boundary conditions
at $x=0$ and $x=L$.  From (\ref{lf81}), we infer that the number of
modes is reduced by a factor of $2$ (compared to the case of closed
boundary conditions), but the construction does not change any
observable constructed from $w^{2}$ or any energy, all based on the
symmetry relation (\ref{addcons}). Importantly, the modes have all
mean 0, which is not the case for other basis, e.g.\ when using
anti-periodic functions. Also note, that this ansatz reproduces the
formula in \cite{AntalDrozGyorgyiRacz2002}.  As an interesting
consequence, we observe that the following systems lead to the same
distribution
\begin{itemize}
\item An elastic object with $N$ degrees of freedom, and closed
boundary conditions.
\item An elastic object with $2N$ degrees of freedom, and open
boundary conditions.
\end{itemize}
The simplest example is a 1-dimensional random walk with closed
boundary conditions, and a 2-dimensional random walk with open
boundary conditions, which can be checked
numerically\footnote{W. Krauth and A. Rosso, private communication.}.

\end{appendix}

\begin{acknowledgments}
We are grateful to W.~Krauth and A.~Rosso for an ongoing collaboration
and numerous stimulating discussions, and we thank E.~Br\'ezin, O.~Narayan and
J.~M.~Schwarz for very useful remarks.
\end{acknowledgments}


\begin{thebibliography}{10}

\bibitem{BookYoung}
A.P. Young,
\newblock {\em Spin glasses and random fields}.
\newblock World Scientific, Singapore, 1997.

\bibitem{Kardar1997}
M.~Kardar,
\newblock {\em Nonequilibrium dynamics of interfaces and lines},
\newblock Phys. Rep. {\bf 301} (1998)   85--112.

\bibitem{DSFisher1998}
D.S. Fisher,
\newblock {\em Collective transport in random media: from superconductors to
  earthquakes},
\newblock Phys. Rep. {\bf 301} (1998)   113--150.

\bibitem{BlatterFeigelmanGeshkenbeinLarkinVinokur1994}
G.~Blatter, M.V. {Feigel'man}, V.B. Geshkenbein, A.I. Larkin  and V.M. Vinokur,
\newblock {\em Vortices in high-temperature superconductors},
\newblock Rev. Mod. Phys. {\bf 66} (1994)   1125.

\bibitem{DSFisher1986}
D.S. Fisher,
\newblock {\em Interface fluctuations in disordered systems: {$5-\epsilon$}
  expansion},
\newblock Phys. Rev. Lett. {\bf 56} (1986)   1964--97.

\bibitem{NattermannStepanowTangLeschhorn1992}
T.~Nattermann, S.~Stepanow, L.-H. Tang  and H.~Leschhorn,
\newblock {\em Dynamics of interface depinning in a disordered medium},
\newblock J. Phys. II (France) {\bf 2} (1992)   1483--8.

\bibitem{NarayanDSFisher1992b}
O.~Narayan and D.S. Fisher,
\newblock {\em Critical behavior of sliding charge-density waves in 4- epsilon
  dimensions},
\newblock Phys. Rev. B {\bf 46} (1992)   11520--49.

\bibitem{NarayanDSFisher1993a}
O.~Narayan and D.S. Fisher,
\newblock {\em Threshold critical dynamics of driven interfaces in random
  media},
\newblock Phys. Rev. B {\bf 48} (1993)   7030--42.

\bibitem{ChauveLeDoussalWiese2000a}
P.~Chauve, P.~Le Doussal  and K.J. Wiese,
\newblock {\em Renormalization of pinned elastic systems: How does it work
  beyond one loop?},
\newblock Phys. Rev. Lett. {\bf 86} (2001)   1785--1788,
\newblock cond-mat/{\bf 0006056}.

\bibitem{LeDoussalWieseChauve2002}
P.~Le Doussal, K.J. Wiese  and P.~Chauve,
\newblock {\em 2-loop functional renormalization group analysis of the
  depinning transition},
\newblock Phys. Rev. B {\bf 66} (2002)   174201,
\newblock cond-mat/0205108.

\bibitem{LeDoussalWieseChauve2002a}
P.~Le Doussal, K.~Wiese  and P.~Chauve,
\newblock {\em Two loop {FRG} study of pinned manifolds},
\newblock in preparation.

\bibitem{RossoKrauth2001a}
A.~Rosso and W.~Krauth,
\newblock {\em Monte carlo dynamics of driven flux lines in disordered media},
\newblock cond-mat\slash {\bf 0107527} (2001).

\bibitem{RossoKrauth2001b}
A.~Rosso and W.~Krauth,
\newblock {\em Origin of the roughness exponent in elastic strings at the
  depinning threshold},
\newblock Phys. Rev. Lett. {\bf 87} (2001)   187002.

\bibitem{RossoKrauth2002}
A.~Rosso and W.~Krauth,
\newblock {\em Roughness at the depinning threshold for a long-range elastic
  string},
\newblock Phys. Rev. E {\bf 65} (2002)   025101/1--4.

\bibitem{TheseAlberto}
A.~Rosso,
\newblock D\'epi\'egeage des vari\'et\'es \'elastiques en milieu al\'eatoire,
  These de Doctorat de l'Universit\'e Paris VI, Paris 30 Sept. 2002.

\bibitem{RossoKrauthLeDoussalVannimenusWiese2002}
A.~Rosso, W.~Krauth, P.~Le Doussal, J.~Vannimenus  and K.J. Wiese,
\newblock {\em Universal interface width distributions at the depinning threshold},
\newblock cond-mat\slash {\bf 0301464} (2003).

\bibitem{DSFisher1985}
D.S. Fisher,
\newblock {\em Sliding charge-density waves as a dynamical critical phenomena},
\newblock Phys. Rev. {\bf B 31} (1985)   1396--1427.

\bibitem{Fisher1985b}
DS. Fisher,
\newblock {\em Random fields, random anisotropies, nonlinear sigma models and
  dimensional reduction},
\newblock Phys. Rev. B {\bf 31} (1985)   7233--51.

\bibitem{VannimenusDerrida2001}
J.~Vannimenus and B.~Derrida,
\newblock {\em A solvable model of interface depinning in random media},
\newblock J. Stat. Phys. {\bf 105} (2001)   1--23.

\bibitem{LeDoussalWiese2001}
P.~Le Doussal and K.J. Wiese,
\newblock {\em Functional renormalization group at large {$N$} for random
  manifolds},
\newblock Phys. Rev. Lett. {\bf 89} (2002),
\newblock cond-mat/{\bf 0109204v1}.

\bibitem{TangKardarDhar1995}
L.-H. Tang, M.~Kardar  and D.~Dhar,
\newblock {\em Driven depinning in anisotropic media},
\newblock Phys. Rev. Lett. {\bf 74} (1995)   920--3.

\bibitem{AlbertBarabasiCarleDougherty1998}
R.~Albert, A.-L. Barabasi, N.~Carle  and A.~Dougherty,
\newblock {\em Driven interfaces in disordered media: determination of
  universality classes from experimental data},
\newblock Phys. Rev. Lett. {\bf 81} (1998)   2926--9.

\bibitem{LeDoussalWiese2002a}
P.~Le Doussal and K.J. Wiese,
\newblock {\em Functional renormalization group for anisotropic depinning and
  relation to branching processes},
\newblock Phys. Rev. E {\bf ???} (2003)   ???,
\newblock cond-mat/0208204.

\bibitem{RossoHartmannKrauth2002}
A.~Rosso, A.K. Hartmann  and W.~Krauth,
\newblock {\em Depinning of elastic manifolds},
\newblock cond-mat\slash {\bf 0207288} (2002).

\bibitem{RotersLubeckUsadel2002}
L.~Roters, S.~Lubeck  and K.~D. Usadel,
\newblock {\em Functional renormalization group for anisotropic depinning and
  relation to branching processes},
\newblock cond-mat\slash {\bf 0207137} (2002).

\bibitem{RamanathanFisher1998}
S.~Ramanathan and D.S. Fisher,
\newblock {\em Onset of propagation of planar cracks in heterogeneous media},
\newblock Phys. Rev. B {\bf 58} (1998)   6026--46.

\bibitem{SchwarzFisher2002}
J.~M. Schwarz and Daniel~S. Fisher,
\newblock {\em Depinning with dynamic stress overshoots: A hybrid of critical
  and pseudohysteretic behavior},
\newblock cond-mat\slash {\bf 0204623} (2002).

\bibitem{PlischkeRaczZia1994}
M.~Plischke, Z.~Racz  and R.K.P. Zia,
\newblock {\em Width distribution of curvature-driven interfaces: a study of
  universality},
\newblock Phys. Rev. E {\bf 50} (1994)   3589--93.

\bibitem{FoltinOerdingRaczWorkmanZia1994}
G.~Foltin, K.~Oerding, Z.~Racz, R.I. Workman  and R.K.P. Zia,
\newblock {\em Width distribution for random-walk interfaces},
\newblock Phys. Rev. E {\bf 50} (1994)   R639--42.

\bibitem{LeDoussalMonthus}
P.~Le Doussal and Cecile Monthus,
\newblock Exact solutions for the statistics of extrema of some random 1D
  landscapes, Application to the equilibrium and the dynamics of the toy model,
  cond-mat/0204168.

\bibitem{ChitraGiamarchiLe1999}
R.~Chitra, T.~Giamarchi  and Le~Doussal,
\newblock {\em Disordered periodic systems at the upper critical dimension},
\newblock Phys. Rev. B {\bf 59} (1999)   4058--65.

\bibitem{Zinn}
J.~Zinn-Justin,
\newblock {\em Quantum Field Theory and Critical Phenomena},
\newblock Oxford University Press, Oxford, 1989.

\bibitem{AntalDrozGyorgyiRacz2002}
T.~Antal, M.~Droz, G.~Gyorgyi  and Z.~Racz,
\newblock {\em Roughness distributions for $1/f^\alpha$ signals},
\newblock Phys. Rev. E {\bf 65} (2002)   046140/1--12.

\bibitem{caveat} To be precise, the calculations of this paper assume
that the zero mode $\overline u$ is exactly set to zero. In that case
momentum sums in internal lines (loops) can exclude $q=0$ in any
diagram. (They are excluded by construction in external lines.) This
restriction does not affect the discussion of the GA but is important
for the calculations of $R$ in section \ref{sec4}.  The present
calculation would apply directly to a numerical simulation where
$\sum_x u_x=0$ is enforced in each disorder configuration (pinned zero
mode). We will discuss the precise connection between the boundary
conditions chosen in simulations such as
\cite{RossoKrauthLeDoussalVannimenusWiese2002} and field theoretical
calculations in a forthcomming publication.

\end{thebibliography}

\end{document}